\documentclass[aps,prd,showpacs,eqsecnum,twocolumn,superscriptaddress]
{revtex4-1}
\usepackage{amsmath,amssymb,graphicx,color}
\usepackage{mathrsfs}

\begin{document}


\title{General relativistic viscous hydrodynamics of differentially
  rotating neutron stars}

\author{Masaru Shibata}
\affiliation{Center of Gravitational Physics, 
Yukawa Institute for Theoretical Physics, 
Kyoto University, Kyoto, 606-8502, Japan} 

\author{Kenta Kiuchi}
\affiliation{Center of Gravitational Physics, 
Yukawa Institute for Theoretical Physics, 
Kyoto University, Kyoto, 606-8502, Japan} 

\author{Yu-ichiro Sekiguchi}
\affiliation{Department of Physics, Toho University, Funabashi, 
Chiba 274-8510, Japan}


\date{\today}
\newcommand{\beq}{\begin{equation}}
\newcommand{\eeq}{\end{equation}}
\newcommand{\beqn}{\begin{eqnarray}}
\newcommand{\eeqn}{\end{eqnarray}}
\newcommand{\pa}{\partial}
\newcommand{\vp}{\varphi}
\newcommand{\varep}{\varepsilon}
\newcommand{\ep}{\epsilon}
\newcommand{\comp}{(M/R)_\infty}
\begin{abstract}
Employing a simplified version of the Israel-Stewart formalism for
general-relativistic shear-viscous hydrodynamics, we perform
axisymmetric general-relativistic simulations for a rotating neutron
star surrounded by a massive torus, which can be formed from
differentially rotating stars.  We show that with our choice of a 
shear-viscous hydrodynamics formalism, the simulations can be stably
performed for a long time scale. We also demonstrate that with a
possibly high shear-viscous coefficient, not only viscous angular
momentum transport works but also an outflow could be driven from a
hot envelope around the neutron star for a time scale $\agt 100$\,ms
with the ejecta mass $\agt 10^{-2}M_\odot$ which is comparable to the
typical mass for dynamical ejecta of binary neutron star mergers.
This suggests that massive neutron stars surrounded by a massive
torus, which are typical outcomes formed after the merger of binary
neutron stars, could be the dominant source for providing neutron-rich
ejecta, if the effective shear viscosity is sufficiently high, i.e.,
if the viscous $\alpha$ parameter is $\agt 10^{-2}$. The present
numerical result indicates the importance of a future high-resolution
magnetohydrodynamics simulation that is the unique approach to clarify
the viscous effect in the merger remnants of binary neutron stars by
the first-principle manner.
\end{abstract}


\maketitle

\section{Introduction}

The recent discoveries of two-solar mass neutron stars~\cite{twosolar}
imply that the equation of state of neutron stars has to be stiff
enough to support the self-gravity of the neutron stars with mass
$\agt 2M_\odot$. Numerical-relativity simulations with stiff equations
of state that provides the maximum neutron-star mass larger than
$2M_\odot$ have shown that massive neutron stars surrounded by a
massive torus are likely to be the canonical remnants formed after the
merger of binary neutron stars of typical total mass
2.6--$2.7M_\odot$~(see, e.g., Refs.~\cite{STU2005,Hotokezaka2013}).
Because a shear layer is inevitably formed on the contact surface of
two neutron stars at the onset of the merger, the Kelvin-Helmholtz
instability~\cite{PR,kiuchi} is activated and the resulting vortex
motion is likely to quickly amplify the magnetic-field strength toward
$\agt 10^{16}$\,G. In addition, because the remnant massive neutron
stars and a torus surrounding them are in general differentially
rotating and magnetized, they shall be subject to magnetorotational
instability (MRI)~\cite{BH98}. As shown by a number of high-resolution
magnetohydrodynamics (MHD) simulations for accretion disks (see, e.g.,
Refs.~\cite{alphamodel,suzuki,local}), MHD turbulence is likely to be
induced for differentially rotating systems and its effect will
determine the subsequent evolution of the system (but see also
Ref.~\cite{GBJJ} for a possibly significant role of neutrinos for
relatively low-magnetic field cases).  As a result, (i) angular
momentum is likely to be transported outward and thermal energy will
be generated by dissipating rotational kinetic energy in the massive
neutron star and surrounding torus and (ii) a massive hot torus is
likely to be further developed around the massive neutron stars.

For exploring MHD processes and resulting turbulent state for the
remnants of binary neutron star mergers, non-axisymmetric
(extremely) high-resolution simulation is necessary if we rely
entirely on a MHD simulation~(see, e.g., Ref.~\cite{kiuchi} for an
effort on this).  The reasons for this are that the wavelength for the
fastest growing modes of the Kelvin-Helmholtz instability and MRI is
much shorter than the stellar size for the typical magnetic-field
strength ($\sim 10^{11}$--$10^{13}$\,G), and in addition, the MHD
turbulence is preserved only in a non-axisymmetric environment: Here,
note that in axisymmetric systems, the turbulence is not preserved for
a long time scale according to the anti-dynamo theorem~\cite{anti}.
This implies that we would need a huge computational cost for studying
realistic evolution of the merger remnants of binary neutron stars
(see, e.g., Ref.~\cite{kiuchi}), and it is practically not an easy
task to obtain a comprehensive picture for the evolution of this
system by systematically performing a large number of MHD simulations
changing neutron-star models and the magnetic-field profiles.  To
date, this problem has not been solved because the well-resolved MHD
simulation has not been done yet.

One phenomenological approach for exploring the evolution of
differentially rotating systems such as the merger remnants is to
employ viscous hydrodynamics in general relativity~\cite{DLSS04}. The
global-scale viscosity is likely to be effectively generated through
the development of the turbulent state induced by the local MHD
processes, and thus, relying on the viscous hydrodynamics implies that
we employ a phenomenological approach, averaging (coarse graining) the
local MHD and turbulence processes. A demerit in this approach is that
we have to artificially input the viscous coefficient, which would be
naturally determined in the MHD simulations. Thus, we cannot obtain
the real answer by one simulation in this approach. We can at best
obtain answers for given values of the viscous coefficient, which has
to be varied for a wide range to obtain a possible variety of the
answers. However, we also have several merits in this approach.
First, we may perform an axisymmetric simulation to follow the
long-term transport processes. We also would not need extremely
high-resolution simulations in this approach, because we do not have
to consider short-wavelength MHD instabilities.  Thus, we can reduce
the computational costs significantly, and hence, with relatively
small computational costs, we are able to systematically explore the
phenomenological evolution of differentially rotating systems
including differentially rotating neutron stars, a torus surrounding
them, and black hole-torus systems.

One caveat for employing viscous hydrodynamics in relativity is that
it could violate the causality if we choose an inappropriate set of
the basic equations. Indeed, in relativistic Navier-Stokes-type
equations~\cite{DLSS04,LL59} in which basic equations are
parabolic-type, the causality is violated. On the other hand, if we
employ Israel-Stewart-type formulations~\cite{Israel1976}, the
resulting equation is not parabolic-type but telegraph-type, and
hence, the causality is preserved~\cite{hiscock83}. In this paper, we employ a
simplified version of the Israel-Stewart formulation to incorporate
shear-viscosity effects neglecting the bulk viscosity and other
transport processes. It is shown that in this case, the hydrodynamics
equations are significantly simplified and they can be numerically
solved in a method quite similar to those for pure hydrodynamics,
while the major effects of the shear viscosity can be qualitatively
captured.

The primary purpose of this paper is to show that our choice of
viscous hydrodynamics formalism works well for long-term simulations
of differentially rotating systems. We perform simulations for
rotating neutron stars surrounded by a torus for a long time scale,
focusing in particular on the long-term mass ejection process from the
torus. In this paper, we do not take into account detailed
microphysics effects such as neutrino transport and we focus only on
the purely viscous hydrodynamics. We plan to present the results of
more detailed studies incorporating microphysics effects in the future
publication.

This paper is organized as follows: In Sec.~II, we describe our
formulation for simplified shear-viscous hydrodynamics.  In
Sec.~III, we apply our formulation to an axisymmetric
general-relativistic simulation for a differentially rotating neutron
star, and show that with a plausible shear viscosity, an outflow may
be driven from a massive neutron star and a torus surrounding it
that are the typical outcomes of binary neutron star
mergers. Section~IV is devoted to a summary.  Throughout this paper,
we employ the units of $c=1=G$ where $c$ and $G$ are the speed of
light and gravitational constant, respectively.

\section{Formulation}

\subsection{Viscous hydrodynamics for general case}

We write the stress-energy tensor of viscous fluid as
\beqn T_{ab}=\rho h u_a u_b +P
g_{ab}- \rho h \nu \tau_{ab}^0\, , \eeqn
where $\rho$ is the rest-mass density, $h$ is the specific enthalpy,
$u^a$ is the four velocity, $P$ is the pressure, $g_{ab}$ is the
spacetime metric, $\nu$ is the viscous coefficient for the shear
stress, and $\tau^0_{ab}$ is the viscous tensor. In terms of the
specific energy $\varep$ and pressure $P$, $h$ is written as
$h=1+\varep+P/\rho$.  $\tau_{ab}^0$ is a symmetric tensor and
satisfies $\tau_{ab}^0 u^a=0$. We suppose that $\nu$ is a function of
$\rho$, $\varep$, and $P$ and will give the relation below.

Taking into account the prescription of Ref.~\cite{Israel1976}, 
we assume that $\tau_{ab}^0$ obeys the following evolution equation:
\beqn
{\cal{L}}_u \tau_{ab}^0=-\zeta (\tau_{ab}^0-\sigma_{ab}), \label{eq2.2}
\eeqn
where ${\cal{L}}_u$ denotes the Lie derivative with respect to $u^a$,
and we set $\sigma_{ab}$ as
\beqn
\sigma_{ab}:=h_a^{~c} h_b^{~d} (\nabla_c u_d + \nabla_d u_c)={\cal{L}}_u h_{ab},
\label{eq:tau}
\eeqn
with $h_{ab}=g_{ab}+u_a u_b$ and $\nabla_a$ the covariant derivative
associated with $g_{ab}$. By introducing Eq.~(\ref{eq2.2}), the
viscous hydrodynamics equation becomes a telegraph-type
equation~\cite{Israel1976,hiscock83}.  Here, $\zeta$ is a non-zero
constant of (time)$^{-1}$ dimension and it has to be chosen in an
appropriate manner so as for $\tau_{ab}^0$ to approach $\sigma_{ab}$
in a short time scale because it is reasonable to suppose that
$\tau_{ab}^0$ should approach $\sigma_{ab}$ in a microphysical
time scale. Thus, we typically choose it so that $\zeta^{-1}$ is
shorter than the dynamical time scale of given systems (but it should
be much longer than the time-step interval of numerical simulations,
$\varDelta t$, in the practical computation).

Equation~(\ref{eq2.2}) can be rewritten as
\beqn
{\cal{L}}_u \tau_{ab}=-\zeta \tau_{ab}^0, \label{eq:tauab}
\eeqn
where $\tau_{ab}:=\tau_{ab}^0 - \zeta h_{ab}$. We employ this equation
for $\tau_{ab}$ as one of the basic equations of viscous
hydrodynamics, and hence, the stress-energy tensor is rewritten as
follows:
\beqn
T_{ab}=\rho h (1-\nu\zeta) u_a u_b  + (P -\rho h \nu \zeta)g_{ab} 
- \rho h \nu \tau_{ab}\, . \nonumber \\
\eeqn

Using the timelike unit vector field normal to spatial
hypersurfaces, $n^a$, and the induced metric on the spatial
hypersurfaces $\gamma_{ab}:=g_{ab}+n_an_b$, we define
\beqn
\rho_{\rm h}&:=&T_{ab} n^a n^b, \\
J_i&:=&-T_{ab} n^a \gamma^b_{~i}\, ,\\
S_{ij}&:=&T_{ab} \gamma^{a}_{~i} \gamma^b_{~j}\, .
\eeqn
Here, the time and spatial components of $n^a$ are written as
$n^{\mu}=(\alpha^{-1},-\alpha^{-1}\beta^i)$ where $\alpha$ and
$\beta^i$ are the lapse function and the shift vector,
respectively. The explicit forms of $\rho_{\rm h}$ and $J_i$ are
\beqn
\rho_{\rm h} &=& \rho h w^2 (1-\nu\zeta)-(P-\rho h \nu \zeta) 
\nonumber \\
&&-\rho h \nu w^{-2} \tau_{ij} \bar u^i \bar u^i\, , \\
J_k&=&\rho h w u_k (1 - \nu\zeta) -\rho h w^{-1} \nu \bar \tau_{k}^{~l} u_l\, ,
\eeqn
where $w:=-n_a u^a=\alpha u^t$, $\bar u^i=\gamma^{ij} u_j$, and $\bar
\tau_{k}^{~l}=\tau_{kj} \gamma^{jl}$ with the bars denoting spatial
components.  Note that we used $\tau_{ab}u^a=0$ and
$\tau_{ab}n^b=\tau_{a}^{~i}u_i w^{-1}$.  We also note that $u_j$ is
equal to $\gamma_{ja} u^a$.

Then, a general-relativistic Navier-Stokes-type equation, derived
from $\gamma_k^{~a} \nabla_b T^b_{a}=0$, is written in a form as
\beqn
&&\pa_t (\sqrt{\gamma} J_k) + \pa_j [\sqrt{\gamma}(\alpha S^j_{~k}
-\beta^j J_k)] \nonumber \\
&&=\sqrt{\gamma}\left(-\rho_{\rm h} \pa_k \alpha +
J_j \pa_k \beta^j -{\alpha \over 2}S_{ij} \pa_k \gamma^{ij}\right),
\eeqn
and the energy equation, derived from $n^a \nabla_b T^b_{~a}=0$, 
is
\beqn
&&\pa_t (\sqrt{\gamma} \rho_{\rm h}) + \pa_j [\sqrt{\gamma}(\alpha J^j
-\beta^j \rho_{\rm h})] \nonumber \\
&&=\sqrt{\gamma}\left(\alpha S_{ij} K^{ij} -J_i D^i \alpha \right),
\eeqn
where $J^i=\gamma^{ij} J_j$, $D_i$ is the covariant derivative
associated with $\gamma_{ij}$, and $K_{ij}$ is the extrinsic curvature
of spatial hypersurfaces.  The transport terms are rewritten using
\beqn
\alpha S^j_{~k}-\beta^j J_k&=&
J_k v^j + \alpha P_{\rm tot} \delta^j_{~k} \nonumber \\
&&-\rho w h \nu \left[{\bar\tau_{k}^{~j} \over u^t}
-{\beta^j + v^j \over w^2}\bar \tau_k^{~l} u_l\right],\\
-\rho_{\rm h}\beta^i + \alpha J^i&=&\rho_{\rm h} v^i + (v^i+\beta^i)P_{\rm tot}
-\rho \alpha h\nu w^{-1}\bar \tau^{ij}u_j \nonumber \\
&&+\rho h \nu w^{-2} (v^i+\beta^i) \bar\tau^{jk}u_j u_k \, ,
\eeqn
where $P_{\rm tot}:=P-\rho h \nu\zeta$ and
$\bar\tau^{jk}=\gamma^{ij}\bar\tau_i^{~k}$. 

In addition to these equations, we have the continuity equation for
the rest-mass density, $\nabla_a (\rho u^a)=0$, which is written as 
usual as
\beqn
\pa_t (\rho \sqrt{\gamma} w) + \pa_j (\rho \sqrt{\gamma} w v^j)=0,
\label{eq:cont}
\eeqn
where $v^j:=u^j/u^t$. 

In viscous hydrodynamics simulations, $\rho_{\rm h}$, $J_i$, and $\rho w$
are evolved (here $\sqrt{\gamma}$ is supposed to be obtained by
solving Einstein's evolution equations).  This implies that it is
straightforward to obtain the following quantities
\beqn
e&:=&{\rho_{\rm h} \over \rho w} 
= h w (1 -\nu\zeta)-{P-\rho h \nu \zeta \over \rho w} \nonumber \\
&& \hskip 2.8cm -h \nu w^{-3} \tau_{ij} \bar u^i\bar u^j\, ,
\label{eq:e}\\
q_j&:=&{J_j \over \rho w}
= h\left[ u_j (1-\nu\zeta) - \nu w^{-2} \bar\tau_j^{~k} u_k \right]. 
\label{eq:qi}
\eeqn
By contrast, $h$ and $w$ have to be calculated by using 
the normalization relation $u^a u_a=-1$, which is written as
\beqn
w^2=\gamma^{ij} u_i u_j + 1. \label{eq:w}
\eeqn
The procedure for a solution of $h$ and $w$ will be described later. 

Spatial components of Eq.~(\ref{eq:tauab}) are explicitly written in
the form
\beqn
u^\mu \pa_\mu \tau_{ij} + \tau_{i \mu} \pa_j u^\mu 
+ \tau_{j\mu} \pa_i u^\mu =-\zeta \tau_{ij}^0\, . 
\eeqn
Here, we focus only on the spatial components of this equation,
because other components of $\tau_{\mu\nu}$ are determined from
$\tau_{ab}u^b=0$.  Multiplying $\rho \alpha \sqrt{\gamma}$ and using the
continuity equation~(\ref{eq:cont}) and $\tau_{\alpha\beta}
u^\beta=0$, we obtain
\beqn
&&\pa_t (\rho w \sqrt{\gamma} \tau_{ij})
+\pa_k \left(\rho w \sqrt{\gamma} \tau_{ij} v^k\right)
\nonumber \\
&&~~
+\rho w \sqrt{\gamma} \left(\tau_{ik}\pa_j v^k + \tau_{jk}\pa_i v^k\right)
=-\rho \alpha \sqrt{\gamma} \zeta \tau_{ij}^0\, .~~~~
\eeqn
Since $\rho w \sqrt{\gamma}$ is determined by solving 
the continuity equation, $\tau_{ij}$ is obtained 
by solving this equation. 

Next we describe how to determine $h$ and $w$. These quantities are
determined from Eqs.~(\ref{eq:e}), (\ref{eq:qi}), and (\ref{eq:w}). 
First, we write Eq.~(\ref{eq:qi}) as
\beqn
q_j=h A_j^{~k} u_k, \label{eq:qj}
\eeqn
where $A_j^{~k}$ is a matrix and a function only of $w^2$ 
because $\bar\tau_j^{~k}$ is obtained by solving the evolution 
equation of $\tau_{ij}$. This implies that by inverting 
Eq.~(\ref{eq:qj}), $u_k$ is written as
\beqn
u_k = h^{-1} (A^{-1})_k^{~j} q_j=: h^{-1} Q_k,\label{eq:u}
\eeqn
and hence, for a given set of $q_j$ and $\tau_{ij}$, $u_k$ can be
considered as a function of $h^{-1}$ and $w^2$.  Substituting
Eq.~(\ref{eq:u}) into (\ref{eq:w}), we obtain a relation between $h$
and $w$ as
\beqn
w^2=h^{-2} \gamma^{ij} Q_i Q_j + 1, \label{eq:Q}
\eeqn
where $Q_k$ can be considered as a function of $w^2$. 

Equation~(\ref{eq:e}) can be also considered as the other relation
between $h$ and $w$ for a given equation of state, $P=P(\rho, \varep)$
or $P=P(\rho, h)$. Thus, by solving simultaneous equations composed of
Eqs.~(\ref{eq:e}) and~(\ref{eq:Q}), we can determine $h$ and $w$.

\subsection{Setting viscous parameter}

In the so-called $\alpha$-viscous model, we have the relation 
(see, e.g., Ref.~\cite{ST})
\beqn
\rho h \nu \Omega \approx \alpha_v P, \label{eq:vis}
\eeqn
where $\Omega$ denotes the local value of the angular velocity and
$\alpha_v$ is the so-called $\alpha$-viscous parameter, which is a
dimensionless constant. In the $\alpha$-viscous model, we assume that
the fluid is in a turbulent state and $\nu$ is written effectively as
$l_{\rm turb}v_{\rm turb}$ where $l_{\rm turb}$ is the size of the
largest turbulent cells and $v_{\rm turb}$ is the velocity of the
turbulent motion relative to the mean gas motion. Since $l_{\rm turb}
< R$ and $v_{\rm turb} < c_s$ where $R$ is the maximum size of the
object concerned (i.e., here the equatorial stellar radius) and $c_s$
is the sound velocity, $\nu$ may be written as $\nu=\alpha_v R c_s$
where $\alpha_v <1$. For rapidly rotating systems, $R\Omega \sim c_s$.
With the definition of the sound velocity, $P/\rho h \sim c_s^2$,
Eq.~(\ref{eq:vis}) is obtained.  We suppose that $\alpha_v$ should be
of the order $10^{-2}$ taking into account the latest results of
high-resolution MHD simulations for accretion disks~(e.g.,
Refs.~\cite{alphamodel,suzuki,local}). 

Thus, in the $\alpha$-viscous model, $\nu$ is written as
\beqn
\nu=\alpha_v c_s^2 \Omega^{-1}. \label{nunu0}
\eeqn
In this paper, we consider viscous hydrodynamics evolution of a
differentially rotating neutron star.  In practice, it is not easy to
appropriately determine $\Omega$ from the local angular velocity of
neutron stars in a dynamical state, and hence, in this work, we simply
set
\beqn
\nu=\alpha_v c_s^2 \Omega_e^{-1},\label{nunu}
\eeqn
where $\Omega_e$ is the angular velocity at the equatorial stellar
surface of the initial state of neutron stars (see, e.g., a filled
circle in the left panel of Fig.~\ref{fig1}).  The relation, $\Omega
\alt \Omega_e$, is satisfied for rapidly rotating neutron stars
and tori (or disks) located in the vicinity
of the neutron stars. Thus, Eq.~(\ref{nunu}) agrees approximately
with Eq.~(\ref{nunu0}) for the outer region and the inner envelope of
the neutron stars, to which we pay special attention in this paper. On
the other hand, $\nu$ is underestimated for a region far from the
rotation axis, to which we do not pay strong attention.

\begin{table*}[t]
\caption{Key quantities for the initial conditions and parameters of
  an equation of state employed in the present numerical simulation:
  Baryon rest mass, $M_*$, gravitational mass, $M$, coordinate
  equatorial radius, $R_e$, circumferential radius at the equatorial
  surface, $R_c$, the maximum rest-mass density, $\rho_{\rm max}$,
  angular velocity at $\varpi=0$, $\Omega_0$, angular velocity at the
  equatorial surface, $\Omega_e$, dimensionless angular momentum,
  $J/M^2$, a polytropic constant, $\kappa_1$, and the value of
  $\rho_1$ (see Eq.~(\ref{eq:EOS})), respectively. We note that the
  initial values of $T_{\rm kin}/M$ and $E_{\rm int}/M$ are $0.048$
  and $0.062$, respectively.  The Kepler angular velocity at the
  equatorial surface is calculated as $\Omega_{\rm
    K}:=\sqrt{M/R_e^3}\approx 9.0 \times 10^3$\,rad/s.
\label{table1}
}
\begin{center}
\begin{tabular}{ccccccccccc} \hline
 ~$M_*\,(M_\odot)$~&~$M\,(M_\odot)$~&~$R_e$\,(km)~&~$R_c$\,(km)~
& $\rho_{\rm max}\,({\rm g/cm^3})$ 
& ~$\Omega_0$\,(rad/s)~ & ~$\Omega_e$\,(rad/s)~ 
& ~~$J/M^2$~~
& $\kappa_1\,({\rm cm^3/s^2g^{1/3}})$ 
& ~$\rho_1\,({\rm g/cm^3})$~ \\
2.64 & 2.37 & 11.7 & 15.7 & $1.00 \times 10^{15}$ & $2.48 \times 10^4$
& $5.25 \times 10^3$ & 0.866 
& $1.24\times 10^{14}$ & $2.04 \times 10^{14}$ \\
 \hline \hline
\end{tabular}
\end{center}
\end{table*}

\subsection{Axisymmetric equations}

We solve viscous hydrodynamics equations in axisymmetric dynamical
spacetime in the following manner.  Einstein's equation in axial
symmetry is solved by a cartoon method~\cite{cartoon,cartoon2}, and
hence, the basic field equations are solved in the $y=0$ plane of
Cartesian coordinates.  Thus, here, we describe viscous hydrodynamics
equations in axial symmetry using Cartesian coordinates with $y=0$.
To do so, the basic equations are first written in cylindrical
coordinates $(\varpi, \varphi, z)$, and then, the coordinate
transformation, $x=\varpi\cos\varphi$ and $y=\varpi\sin\varphi$,
should be carried out. The resulting equations are as follows: The
continuity equation is written as
\beqn
\pa_t \rho_* + {1\over x} \pa_x (\rho_* x v^x)
+\pa_z (\rho_* v^z)=0, \label{eq:cont2d}
\eeqn
where $\rho_*:=\rho w \sqrt{\hat\gamma}$ and
$\hat\gamma=\gamma/\varpi^2$: $\varpi^2$ is the determinant of the
flat-space metric in the cylindrical coordinates.  Each component of
the viscous hydrodynamics equation is written in the forms
\beqn
&&\pa_t S_x + \pa_x \left[S_x v^x + \alpha \sqrt{\hat\gamma} P_{\rm tot}
-\rho_* h \nu \hat \tau_x^{~x} \right]\nonumber \\
&&~~~~~~+\pa_z \left[S_x v^z-\rho_* h \nu \hat \tau_x^{~z}\right] \nonumber \\
&&~~=F_x+{1 \over x}\left[S_y v^y-S_xv^x \right]
-{\rho_*h\nu  \over x} \left[\hat\tau_y^{~y}-\hat\tau_x^{~x}\right],~~~~~
\label{eq:sx2d}\\
&&\pa_t S_z + \pa_x \left[S_z v^x -\rho_* h \nu \hat\tau_z^{~x}\right] 
\nonumber \\
&&~~~~~~+\pa_z \left[S_z v^z + \alpha \sqrt{\hat\gamma} P_{\rm tot}
-\rho_* h \nu \hat\tau_z^{~z} \right] \nonumber \\
&&~~=F_z-{1 \over x}\left[S_z v^x - \rho_*h\nu \hat\tau_z^{~x}\right],
\eeqn
\beqn
&&\pa_t S_y + {1 \over x^2} \pa_x \left[x^2 \left(
S_y v^x -\rho_* h \nu \hat\tau_y^{~x} \right)\right] \nonumber \\
&&~~~~~~+\pa_z \left(S_y v^z -\rho_* h \nu \hat\tau_y^{~z} \right) =0,
\label{eq:ang2d}
\eeqn
where $S_i:=\sqrt{\hat\gamma}\,J_i$, and 
\beqn
\hat\tau_j^{~i}&:=&{\bar\tau_{j}^{~i} \over u^t}
-{\beta^i+v^i \over w^2} \bar\tau_j^{~k}u_k, \nonumber \\
F_p&:=&-S_0\pa_p \alpha + S_i\pa_p \beta^i-{\alpha \over 2} \sqrt{\hat \gamma}
S_{ij}\pa_p\gamma^{ij}.
\eeqn
Here, the index $p$ denotes $x$ or $z$, and $i$, $j$, and $k$ do $x$, $y$,
or $z$.

The energy equation is written in the form
\beqn
&&\pa_t S_0 \nonumber \\
&&~~~+ {1 \over x} \pa_x \left[x \left(
S_0 v^x +(\beta^x+v^x)\sqrt{\hat\gamma}P_{\rm tot}
-{\rho_* h \nu \over w} \hat\tau^{xk}u_k \right)\right] \nonumber \\
&&~~~+\pa_z \left(S_0 v^z+(\beta^z+v^z)\sqrt{\hat\gamma}P_{\rm tot}
-{\rho_* h \nu \over w}\hat\tau^{zk}u_k \right)  \nonumber \\
&&~=\alpha\sqrt{\hat\gamma}\, S_{ij} K^{ij}-S_i D^i \alpha,
\label{eq:so2d}
\eeqn
where $S_0:=\sqrt{\hat \gamma}\rho_{\rm h}$ and
$\hat\tau^{ij}=\gamma^{ik}\hat\tau_k^{~j}$.  We note that the terms
associated with $\hat \tau_{ij}$ in
Eqs.~(\ref{eq:ang2d}) and (\ref{eq:so2d}) are
responsible for the angular-momentum transport and viscous heating,
respectively.

The method for a solution of these hydrodynamics equations is the same
as in Refs.~\cite{cartoon2,SS2012}: The transport terms are
specifically evaluated using a Kurganov-Tadmor scheme~\cite{KT} with a
piecewise parabolic reconstruction for the quantities of cell
interfaces.  We do not take into account the modification of the
characteristic speed by the viscous effect for simplicity because the
local transport time scale of fluid elements, $R/|v^i|$, is much
shorter than the viscous time scale, $R^2/\nu$, in our choice of the
alpha viscosity (here, $R$ denotes a characteristic length scale of
the system). That is, the local characteristic speed of the fluid
dynamics would be modified only slightly by the viscous effect.

The evolution equations for $\tau_{ij}$ are written as
\beqn
&& \pa_t(\rho_* \tau_{xx}) + \pa_x (\rho_* \tau_{xx}v^x) 
+ \pa_z(\rho_*\tau_{xx}v^z) \nonumber \\
&&~~~=-2 \rho_*\biggl(\tau_{xx} \pa_x v^x + \tau_{xy} \pa_x v^y 
+\tau_{xz} \pa_x v^z -\tau_{xy}{v^y \over x}\biggr) \nonumber \\
&&~~~~~-{\zeta \rho_* \over u^t}\tau^0_{xx} 
-\rho_* \tau_{xx}{v^x \over x}, \\
&& \pa_t(\rho_* \tau_{xy}) + \pa_x (\rho_* \tau_{xy}v^x) 
+ \pa_z(\rho_*\tau_{xy}v^z) \nonumber \\
&&~~~=-\rho_*\biggl(\tau_{xy} \pa_x v^x + \tau_{yy} \pa_x v^y 
+\tau_{yz} \pa_x v^z -\tau_{yy}{v^y \over x}\biggr) \nonumber \\
&&~~~~~-{\zeta \rho_* \over u^t}\tau^0_{xy} 
-2 \rho_* \tau_{xy}{v^x \over x},
\eeqn
\beqn
&& \pa_t(\rho_* \tau_{xz}) + \pa_x (\rho_* \tau_{xz}v^x) 
+ \pa_z(\rho_*\tau_{xz}v^z) \nonumber \\
&&~~~=- \rho_*\biggl(\tau_{xx} \pa_z v^x + \tau_{xy} \pa_z v^y 
+\tau_{xz} \pa_z v^z +\tau_{xz}\pa_x v^x   \nonumber \\
&&~~~~~~~~~~+\tau_{yz} \pa_z v^y + \tau_{zz} \pa_x v^z 
-\tau_{yz}{v^y \over x}\biggr) \nonumber \\
&&~~~~~-{\zeta \rho_* \over u^t}\tau^0_{xz} 
-\rho_* \tau_{xz}{v^x \over x},
\\
&& \pa_t(\rho_* \tau_{yy}) + \pa_x (\rho_* \tau_{yy}v^x) 
+ \pa_z(\rho_*\tau_{yy}v^z) \nonumber \\
&&~~~=-{\zeta \rho_* \over u^t}\tau^0_{yy} 
-3\rho_* \tau_{yy}{v^x \over x},
\\
&& \pa_t(\rho_* \tau_{yz}) + \pa_x (\rho_* \tau_{yz}v^x) 
+ \pa_z(\rho_*\tau_{yz}v^z) \nonumber \\
&&~~~=- \rho_*\biggl(\tau_{xy} \pa_z v^x + \tau_{yy} \pa_z v^y 
+\tau_{yz} \pa_z v^z \biggr)   \nonumber \\
&&~~~~~-{\zeta \rho_* \over u^t}\tau^0_{yz} 
-2 \rho_* \tau_{yz}{v^x \over x},
\\
&& \pa_t(\rho_* \tau_{zz}) + \pa_x (\rho_* \tau_{zz}v^x) 
+ \pa_z(\rho_*\tau_{zz}v^z) \nonumber \\
&&~~~=-2 \rho_*\biggl(\tau_{xz} \pa_z v^x + \tau_{yz} \pa_z v^y 
+\tau_{zz} \pa_z v^z \biggr)   \nonumber \\
&&~~~~~-{\zeta \rho_* \over u^t}\tau^0_{zz} 
-\rho_* \tau_{zz}{v^x \over x}.
\eeqn
For the characteristic speed of these equations, we simply employ
$v^x$ and $v^z$ for the $x$- and $z$-directions, respectively. 
From the regularity condition for the tensor quantity, we 
find the boundary conditions for $\tau_{ij}$ at the symmetric 
axis, $\varpi=0$, as $\tau_{xx}=\tau_{yy}$, $\tau_{xy} \propto \varpi^2$, 
$\tau_{xz} \propto \varpi$, and $\tau_{yz} \propto \varpi$. 

From Eqs.~(\ref{eq:cont2d}) and (\ref{eq:ang2d}), it is immediately
found that the baryon rest mass, $M_*$, and angular momentum, $J$,
are conserved quantities, which are defined by
\beqn
&&M_*:=2\pi \int \rho_* x dx dz,\\ 
&&J:=2\pi \int S_y x^2 dx dz. 
\eeqn
In numerical simulations, we monitor these quantities and check that
they are preserved to be (approximately) constant. Note that these
quantities are precisely conserved unless matter is ejected from the
outer boundaries, because we solve the conservative forms for the
equations of $\rho_*$ and $S_y$.  We also monitor the kinetic energy
and internal energy defined, respectively, by
\beqn
T_{\rm kin}&=& \pi \int \rho_* h u_k v^k  x dx dz, \\
E_{\rm int}&=& 2 \pi \int \rho_* \varep  x dx dz. 
\eeqn
These values clearly show how the viscous dissipation proceeds: 
$T_{\rm kin}$ and $T_{\rm kin}/E_{\rm int}$ decrease by converting the
kinetic energy to the internal energy.

We also calculate the mass and energy fluxes through a sphere far from
the central object and evaluate the total mass and energy for the
outflow component.  The mass and energy fluxes are defined,
respectively, by
\beqn
&& F_M= 2\pi \oint_{r={\rm const}} d(\cos\theta)\, \rho_* v^r r^2, \\ 
&& F_E=-2\pi \oint_{r={\rm const}} d(\cos\theta)\, 
T_t^{~r} r^2 \alpha \sqrt{\hat \gamma}\, .
\eeqn
Then, we calculate the outflowed mass and energy as functions of time by
\beqn
&&M_{\rm out}(t)=\int^t_0 F_M dt',\\
&&E_{\rm out}(t)=\int^t_0 F_E dt'. 
\eeqn
Here, the internal energy of the outflow component is much smaller
than the kinetic energy if we evaluate the outflow quantity in a far
zone. Thus, we define the kinetic energy of the outflow by $T_{\rm
  out} \approx E_{\rm out} -M_{\rm out}$.

\begin{figure}[t]
\begin{center}
\includegraphics[width=85mm]{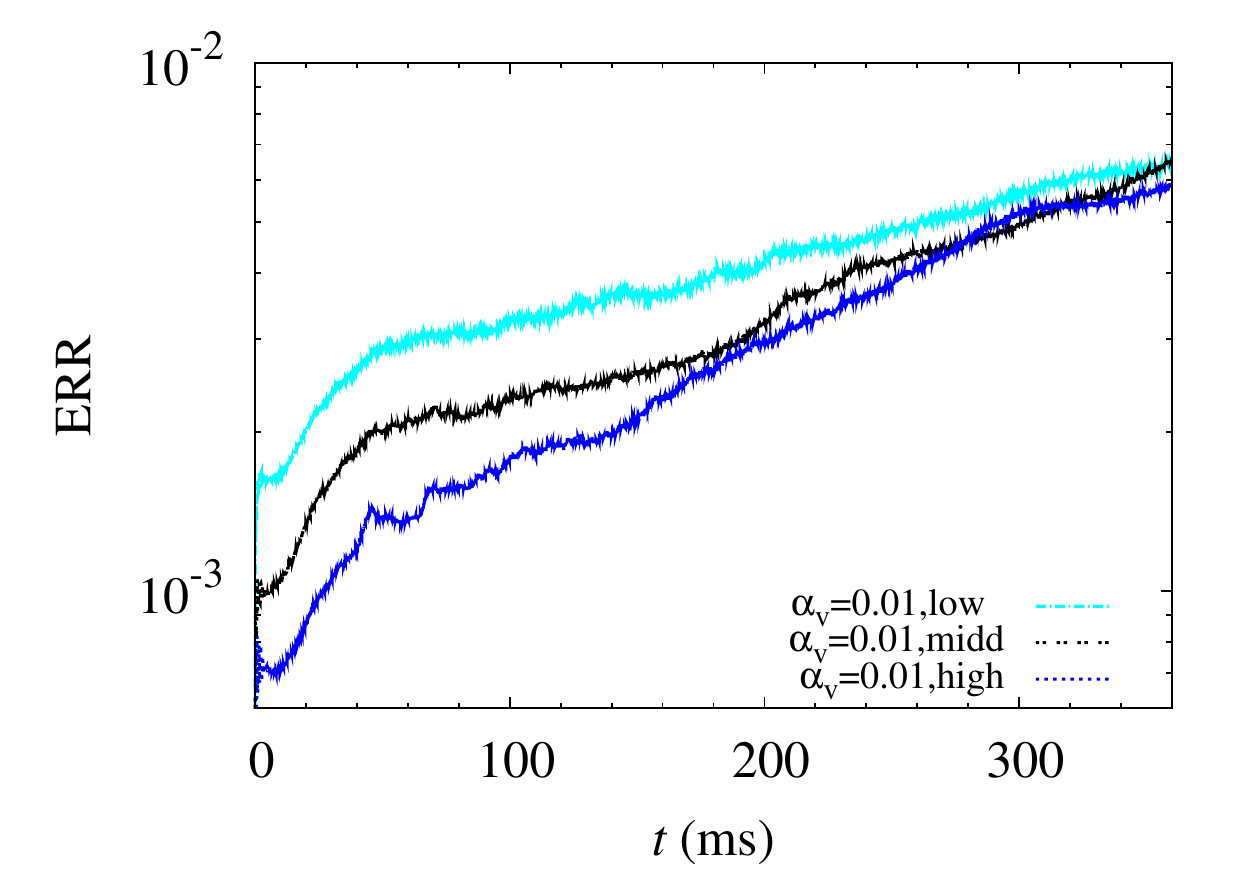}
\caption{Evolution of ERR defined in Eq.~(\ref{eqERR}) for the model
  with $\alpha_v=0.01$. The results with three different grid
  resolutions are plotted.
\label{fig0}}
\end{center}
\end{figure}

Before closing Sec.~II, we should comment on the method of evaluating
the derivative of $v^i$, which appears in the equations for
$\tau_{ij}$ and does not appear in ideal fluid hydrodynamics.  For the
numerical results presented in this paper, we evaluate it by simple
second-order centered finite differencing. However, $v^i$ is not
always continuous and hence this treatment could introduce a
non-convergent error.  We monitor the violation of the Hamiltonian
constraint, $H=0$, in particular focusing on the following
rest-mass-averaged quantity,
\beq
{\rm ERR}={1 \over M_*} \int \rho_* {|H| \over \sum_k |H_k|} d^3x,
\label{eqERR}
\eeq
where $H=\sum_k H_k$ and $H_k$ denotes individual components in $H$
like $16\pi \rho_{\rm h}$, $K_{ij}K^{ij}$, $(K_k^{~k})^2$, and
three-dimensional Ricci scalar.  ERR shows the global violation of the
Hamiltonian constraint. For ERR=0, the constraint is satisfied,
while for ERR=1, the Hamiltonian constraint is by 100\% violated.
Figure~\ref{fig0} shows the evolution of ERR for the model with
$\alpha_v=0.01$ (see the next section for the details of our
models). This figure illustrates that the convergence with respect to
the grid resolution is far less than second order.  However, the
degree of the violation is reasonably small with ERR $\alt 0.01$ in
our simulation time.  This approximately indicates that the
Hamiltonian constraint is satisfied within 1\% error, and hence, we
suppose that the results obtained in this paper would be reliable at
least in our present simulation time. 

For a long-term simulation, however, the violation is accumulated and
eventually it could be so large that we are prohibited to derive a
reliable numerical result or the computation crashes.
For suppressing the numerical error, we will need to implement a
better scheme of evaluating this slowly-convergent derivative term.


\section{Numerical simulation}

\begin{figure*}[t]
\begin{center}
\includegraphics[width=85mm]{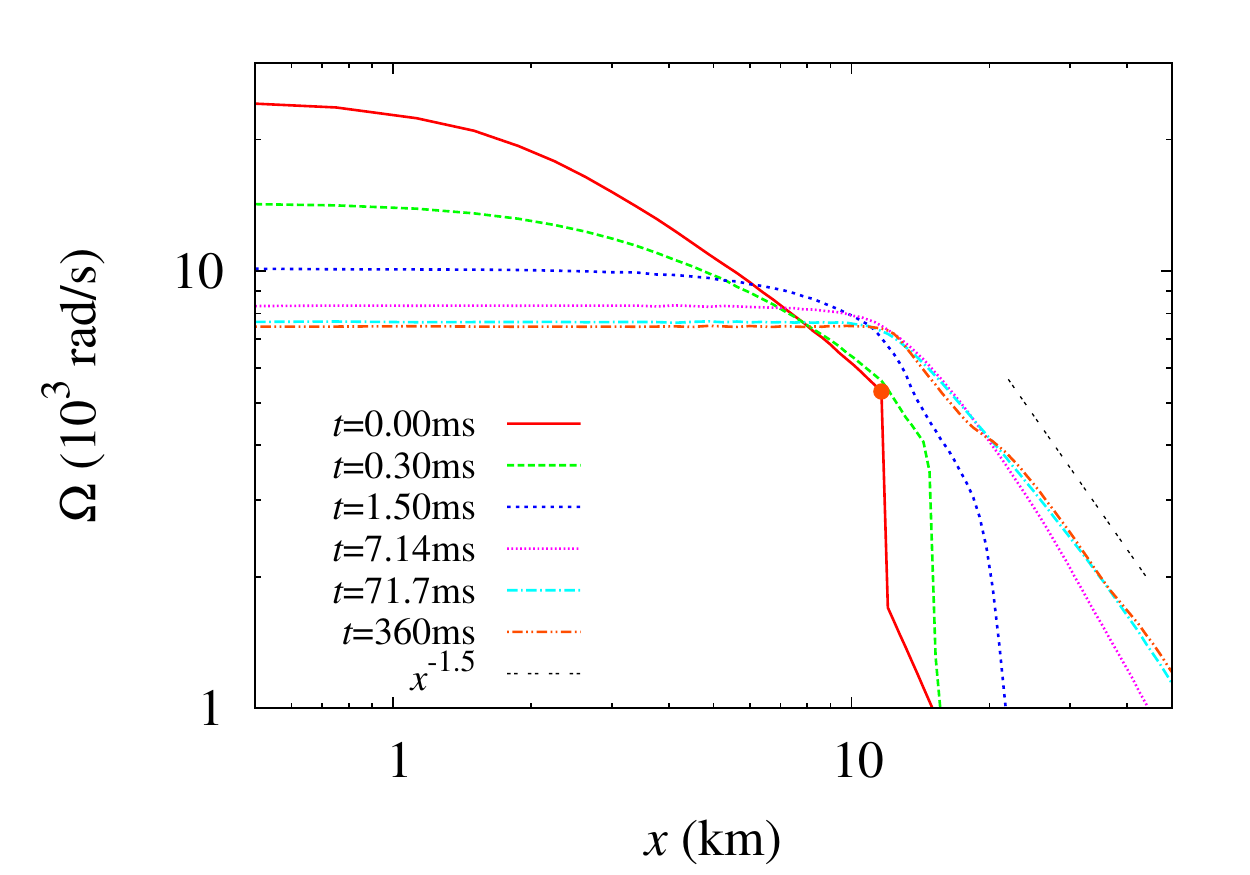}~~~
\includegraphics[width=85mm]{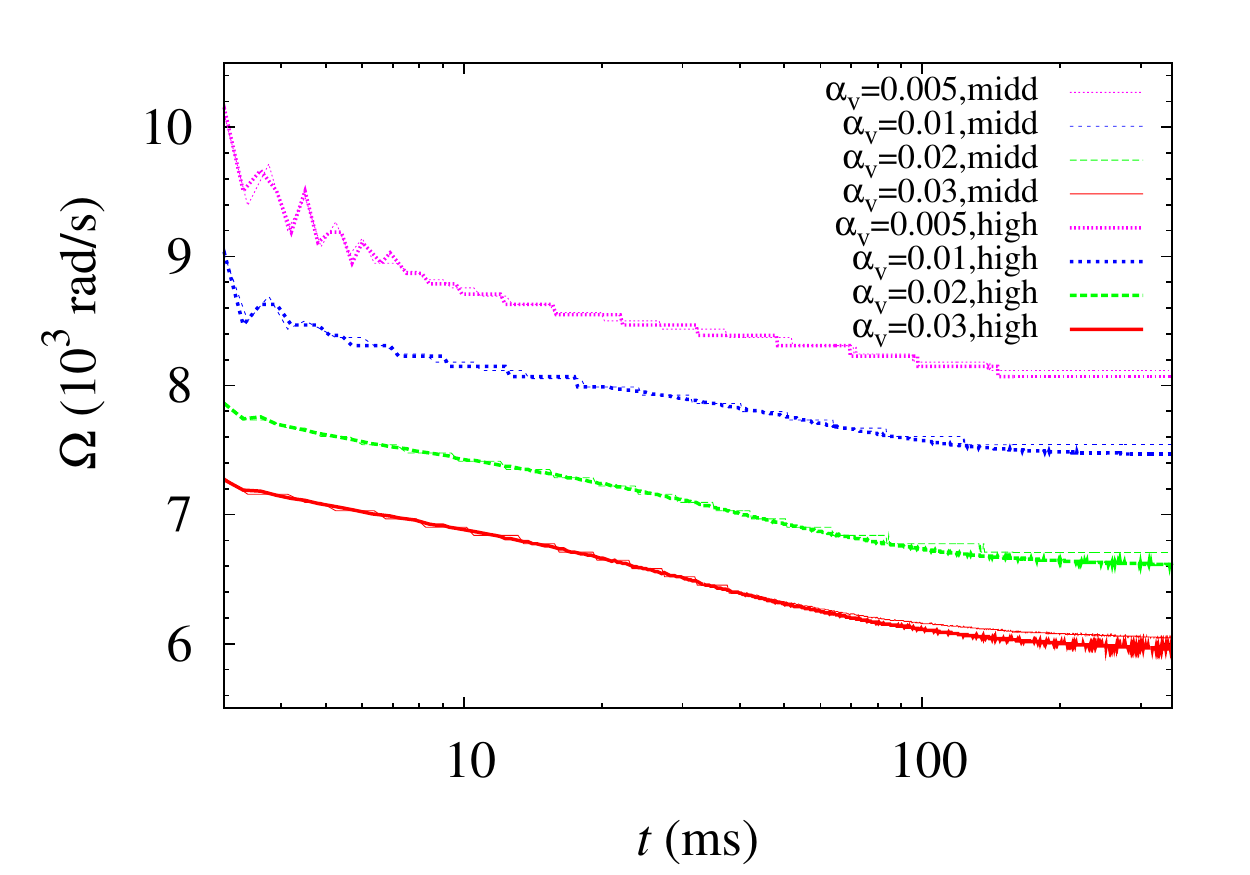}
\caption{Left: Evolution of the profile of $\Omega$ as a function of
  the cylindrical radius one the equatorial plane. For this model,
  $\alpha_v=0.01$.  The dot-dot slope denotes the inclination of
  $x^{-3/2}$. The filled circle denotes $\Omega_e$ at $t=0$. Right:
  Evolution of the angular velocity at the center for
  $\alpha_v=0.005$--0.03 and for the middle and high grid resolutions.
\label{fig1}}
\end{center}
\end{figure*}

\subsection{Brief summary of simulation setting}

Our method for a solution of Einstein's equation is the same as that
in Ref.~\cite{SS2012}: We employ the original version of
Baumgarte-Shapiro-Shibata-Nakamura formulation with a puncture-type 
gauge~\cite{BSSN}.  The gravitational field equations are solved in
the fourth-order finite differencing scheme.  The axial symmetry
is imposed using the cartoon method~\cite{cartoon,cartoon2,SS2012}, as
already mentioned. A fourth-order Lagrange interpolation scheme is
used for implementing the cartoon scheme.

A differentially rotating neutron star, which is used as an initial
condition, is modeled employing a piecewise polytropic equation
of state with two pieces:
\beqn
P_{\rm pwp}=\left\{
\begin{array}{cc}
\kappa_1 \rho^{\Gamma_1} & \rho \leq \rho_1, \\
\kappa_2 \rho^{\Gamma_2} & \rho \geq \rho_1,
\label{eq:EOS}
\end{array}
\right.
\eeqn
where $\kappa_1$ and $\kappa_2$ are polytropic constants and
$\Gamma_1$ and $\Gamma_2$ are polytropic indices, respectively.
$\rho_1$ is a constant of the nuclear-density order: We here set it to
be $\approx 2.0 \times 10^{14}\,{\rm g/cm^3}$.  In this work, we
choose $\Gamma_1=4/3$ and $\Gamma_2=11/4$, respectively.

For constructing initial models, we assume a very simple profile for 
the angular velocity given by $u^t u_\varphi=\hat A^2(\Omega_0-\Omega)$
where $\Omega_0$ is the angular velocity along the rotation axis.  As
in Ref.~\cite{PRL}, we set $\hat A=0.8R_e$, and then, the angular
velocity is approximately given by
\beqn
\Omega \approx {\Omega_0 (0.8R_e)^2 \over \varpi^2 + (0.8R_e)^2},
\eeqn
where $R_e$ denotes the equatorial coordinate stellar radius.  For the
simulation, we pick up a high-mass neutron star with the coordinate
axial ratio 0.3 (i.e., the ratio of the polar coordinate radius to
$R_e$ is 0.3).  The important quantities for the initial condition is
listed in Table~I.

We note that the {\em initial} angular-velocity profile employed in
this paper is qualitatively different from that of the merger remnants
of binary neutron stars: For the realistic binary neutron star merger,
the angular velocity near the rotation axis is rather slow, reflecting
the fact that the velocity vectors of two neutron stars have the
counter direction at the merger, and hence, shocks that dissipate
their kinetic energy are formed~\cite{STU2005}. Then, the merger
remnant neutron star is weakly differentially rotating and surrounded
by a thick torus.  Starting from the differentially rotating neutron
star employed in this paper, we soon have a (approximately) rigidly
rotating neutron star surrounded by a massive torus, as we show below.
Such outcome is similar to the merger remnant. One of the major
purposes of this paper is to pay attention to long-term evolution of
this type of the outcome.

During numerical evolution, we employ a modified version 
of the piecewise polytropic equation of state in the form
\beqn
P=P_{\rm pwp}(\rho)+(\Gamma-1) \rho [\varep-\varep_{\rm pwp}(\rho)],
\eeqn
where $\varep_{\rm pwp}(\rho)$ denotes the specific internal energy
associated with $P_{\rm pwp}$ satisfying $d\varep_{\rm pwp}=-P_{\rm
  pwp}d \rho^{-1}$ and the adiabatic constant $\Gamma$ is set to be
$3/2$.  The second term is added to take into account a shock heating
effect. We choose a relatively small value of $\Gamma$ in this work to
mildly incorporate the shock heating effects. 

Numerical simulations are performed in cylindrical coordinates 
$(x, z)$, and a nonuniform grid is used for $x$ and $z$. 
Specifically, we employ the following grid structure
(the same profile is chosen for $z$)
\beqn
\varDelta x=\left\{
\begin{array}{ll}
\varDelta x_0 &~~~ x \leq x_{\rm in}, \\
\varDelta x_i=f\varDelta x_{i-1} &~~~ x > x_{\rm in},
\end{array}
\right.
\eeqn
where $\varDelta x_0$ is the grid spacing in an inner region with
$x_{\rm in} \approx 1.1R_e$.  $\varDelta x_i:= x_{i+1}- x_{i}$ with
$x_i$ being the location of the $i$-th grid point. At $i={\rm in}$,
$\varDelta x_i=\varDelta x_0$.  $f$ determines the nonuniform degree
of the grid spacing and we set it to be $1.01$.  We change $\varDelta
x_0$ as $R_e/75$ (low resolution), $R_e/100$ (middle resolution), and
$R_e/125$ (high resolution) to confirm that the dependence of the
numerical results on the grid resolution is weak. We note that
$R_e/125=94$\,m in our model.  The outer boundary is located at
$\approx 230R_e \approx 2600$\,km for all the grid resolutions.
Unless otherwise stated, we will show the results in the
high-resolution runs in the following.

\begin{figure*}[t]
\begin{center}
\includegraphics[width=82mm]{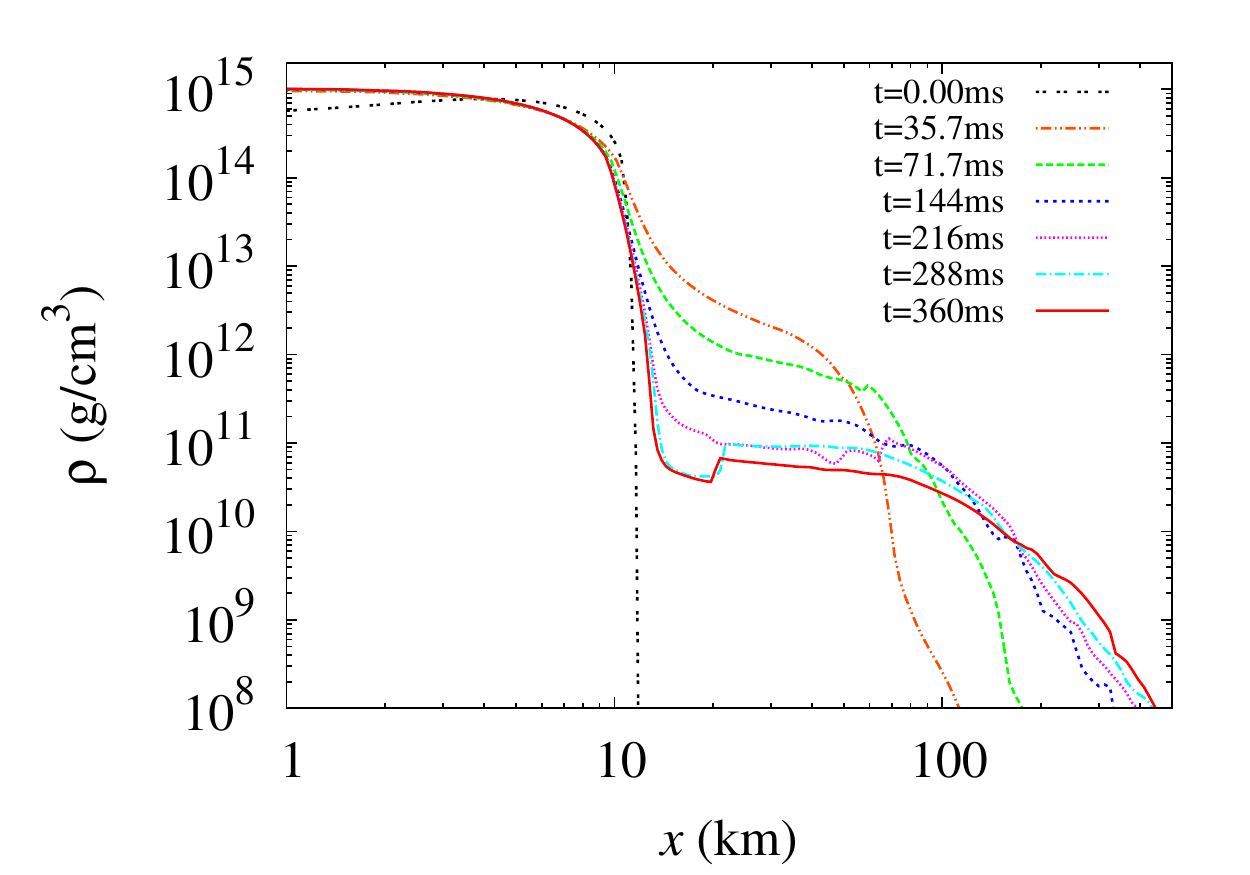}~~~~~
\includegraphics[width=82mm]{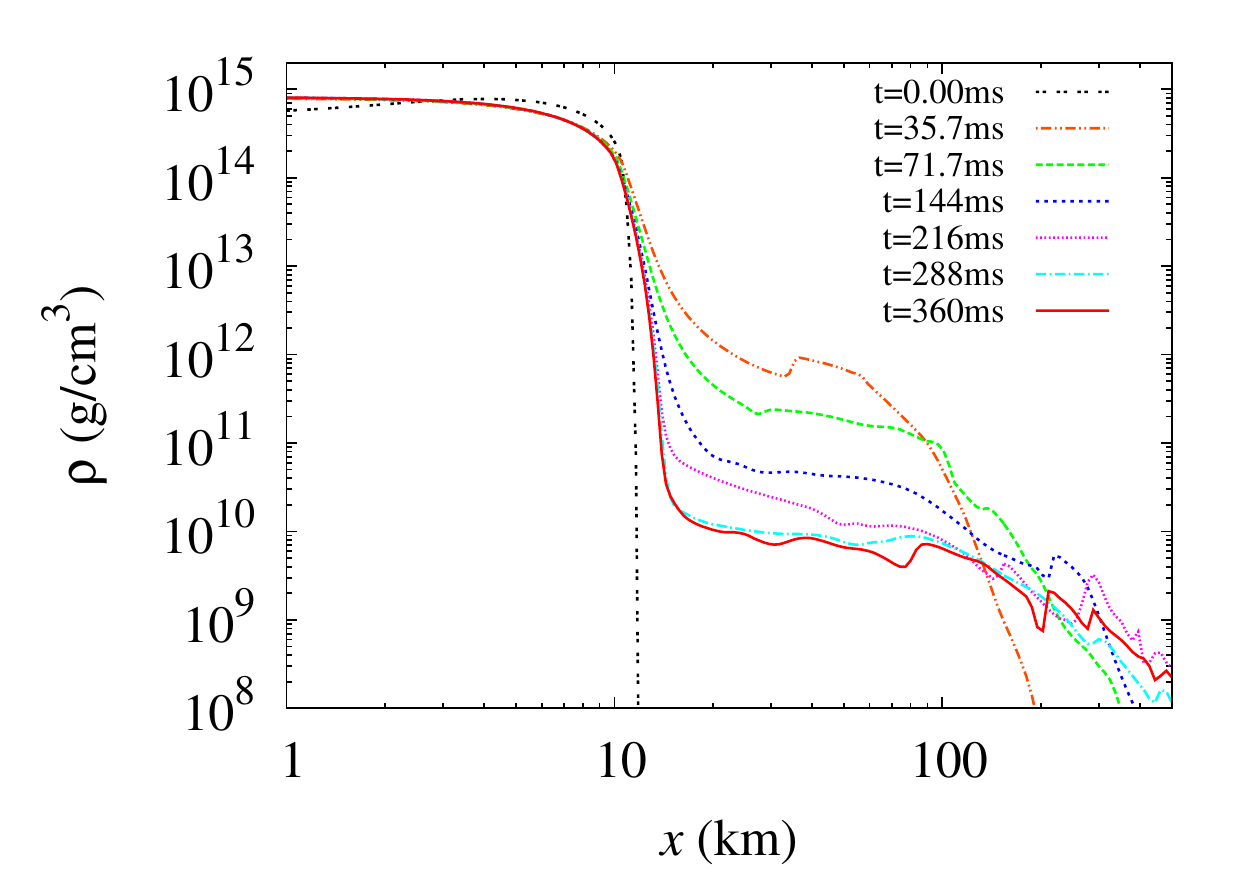}
\caption{Evolution of the density profile on the equatorial plane as a
  function of the cylindrical radius for $\alpha_v=0.01$ (left) and
  0.03 (right).
\label{fig2}}
\end{center}
\end{figure*}

When using a nonuniform grid, numerical instability is often induced
in a long-term simulation due to the gradual growth of high-frequency
noises in the geometric variables, in particular in the extrinsic
curvature. To suppress the growth of unstable modes associated with
the numerical noises, we incorporate a six-order Kreiss-Oliger-type
dissipation term as (see, e.g., Ref.~\cite{bernd})
\beqn
Q \rightarrow Q + \sigma {\varDelta x_0^6 \over 720} Q^{(6)}, \label{KO}
\eeqn
where $\sigma$ is a constant of order unity and $Q$ denotes the 
geometric quantities. $Q^{(6)}$ in the present axisymmetric simulation
is calculated by
\beqn
Q^{(6)}={\pa^6 Q \over \pa x^6} + {\pa^6 Q \over \pa z^6}.
\eeqn
Note that the coefficient in the second term of Eq.~(\ref{KO}) is
written in terms of $\varDelta x_0$ (not $\varDelta x_i$) because the
accumulation of the high-frequency noise causes the problem only in
the inner region.  We find that with this prescription, the ERR in
Eq.~(\ref{eqERR}) can be kept to be $\alt 10^{-2}$ for $t \alt
500$\,ms (see Fig.~\ref{fig0}).

The viscous coefficient is written in the form of Eq.~(\ref{nunu}).
We choose $\alpha_v=0.005$, 0.01, 0.02, and 0.03. $\zeta$ is set to be
$\approx 3\Omega_0$. The viscous angular momentum transport time scale
is approximately defined by $R^2/\nu$~\cite{Kato} and estimated to be
\beqn
t_{\rm vis}&\approx &14\,{\rm ms} \left({\alpha_v \over 0.01}\right)^{-1}
\left({c_s \over 0.2c}\right)^{-2}
\left({R \over 10\,{\rm km}}\right)^{2} \nonumber \\
&&~~~~\times \left({\Omega \over 5 \times 10^3\,{\rm rad/s}}\right),
\label{eq3.5}
\eeqn
where we assumed Eq.~(\ref{nunu0}) for $\nu$. In the vicinity of the
rotation axis (for a small value of $R$ and a high value of $c_s$),
the time scale should be initially short.  Thus, in $\sim 10
(\alpha_v/0.01)^{-1}$\,ms, the angular momentum is expected to be
transported outward in the differentially rotating neutron star
initially prepared.

\subsection{Numerical results}

\begin{figure*}[t]
\begin{center}
\includegraphics[width=56mm]{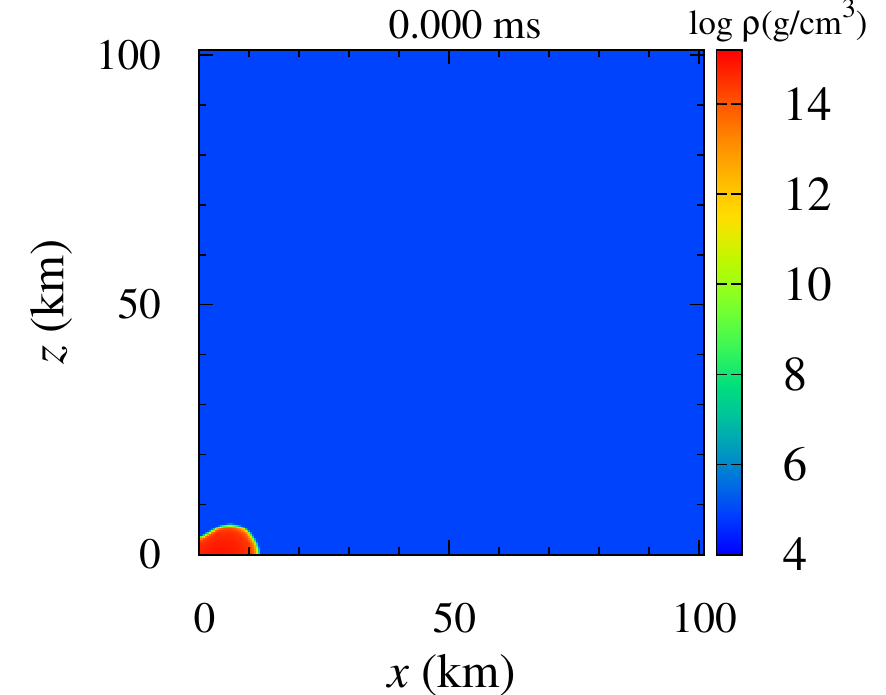}
\includegraphics[width=56mm]{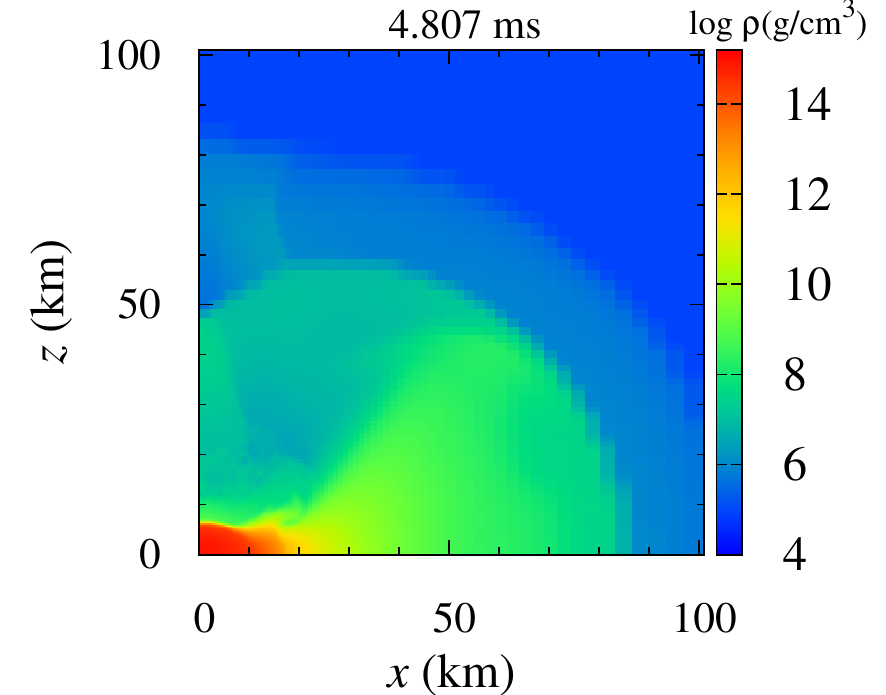}
\includegraphics[width=56mm]{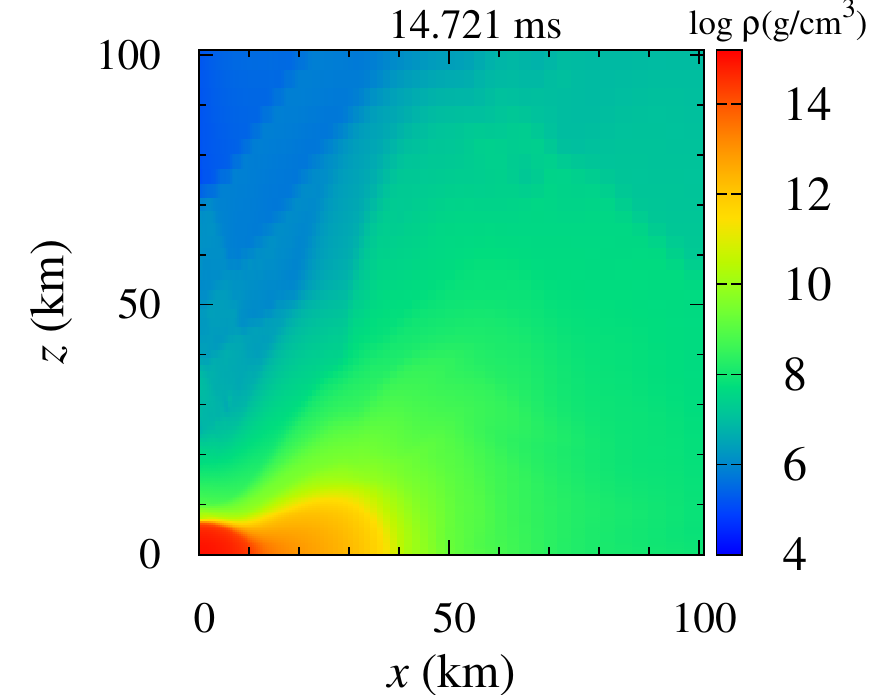} \\
\vspace{0.2cm}
\includegraphics[width=56mm]{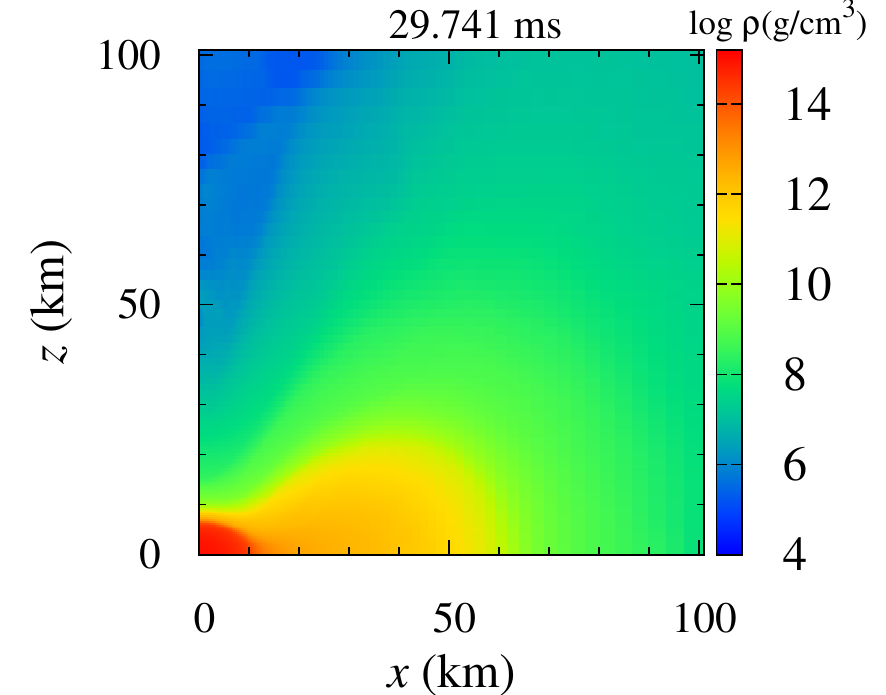}
\includegraphics[width=56mm]{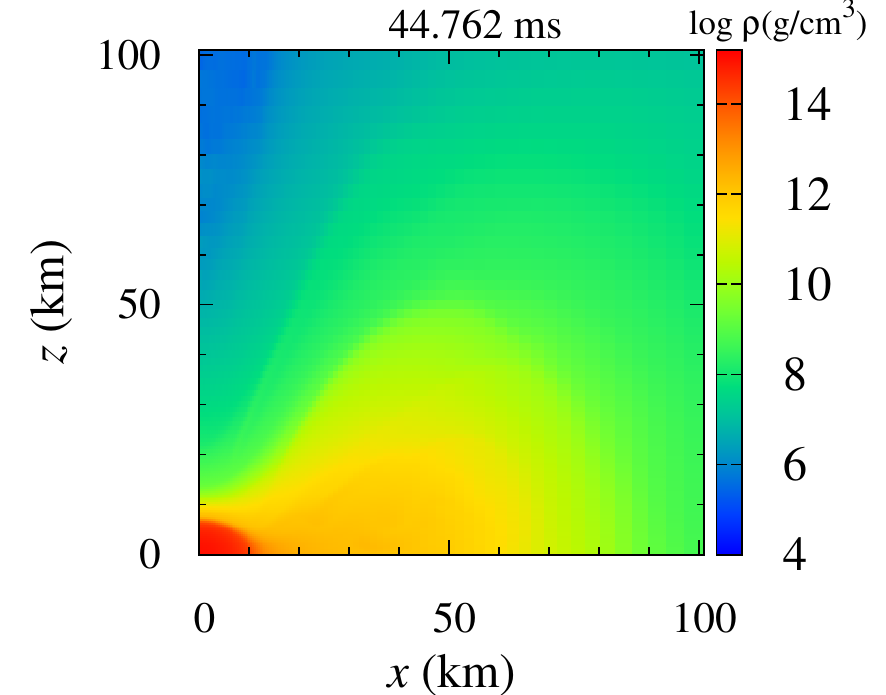}
\includegraphics[width=56mm]{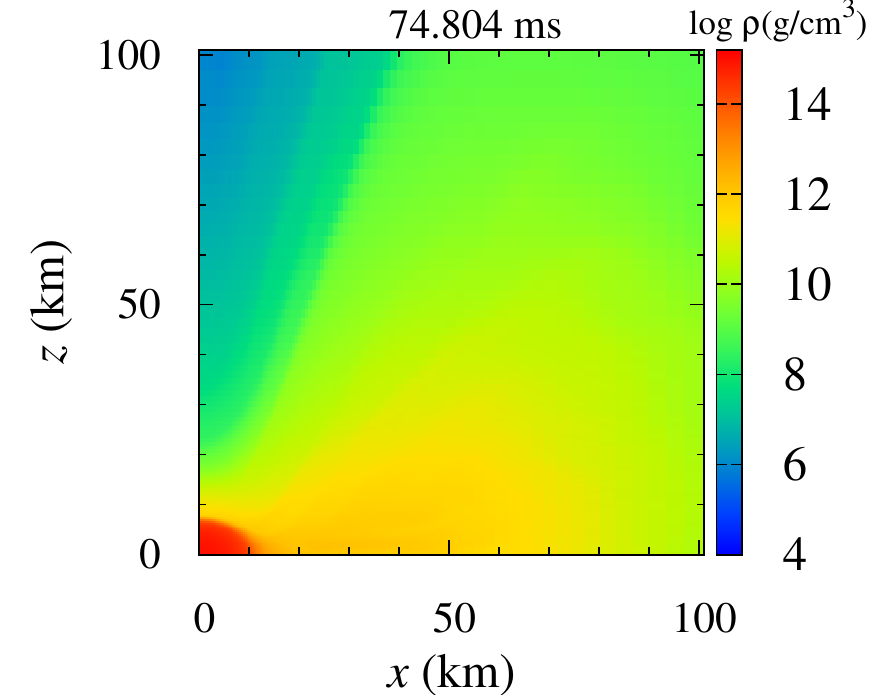} \\
\vspace{0.2cm}
\includegraphics[width=56mm]{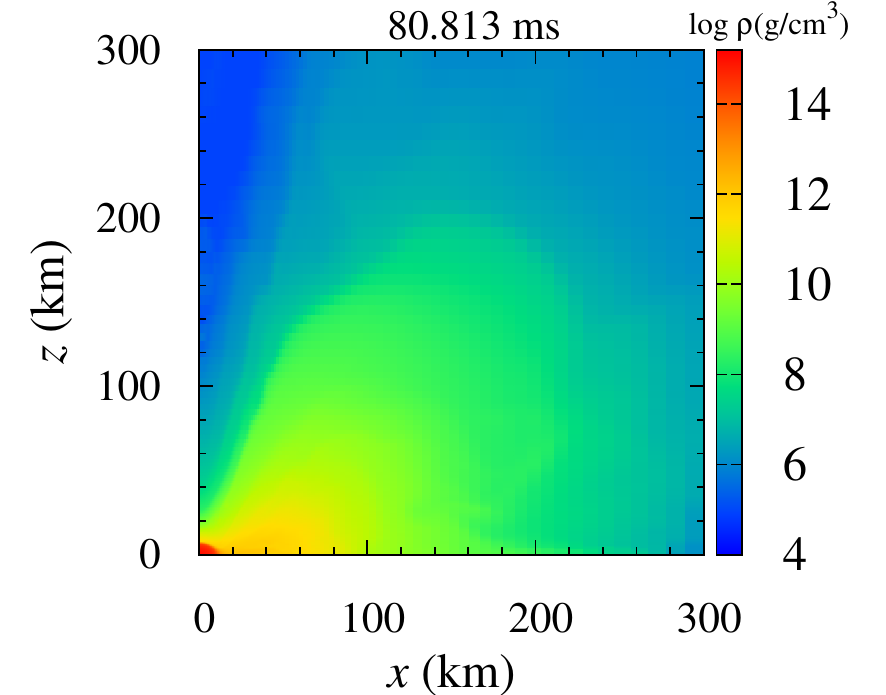}
\includegraphics[width=56mm]{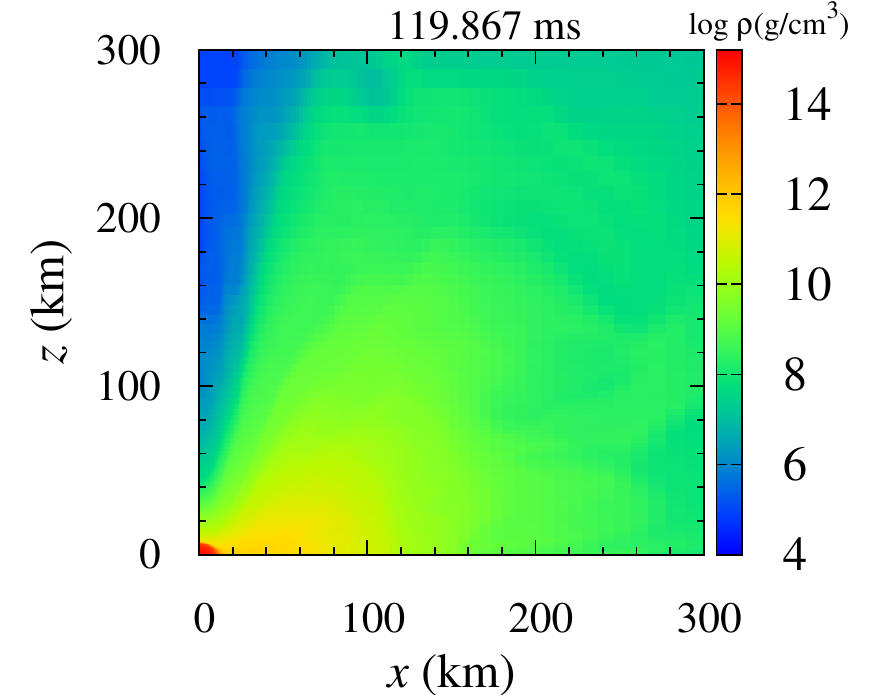}
\includegraphics[width=56mm]{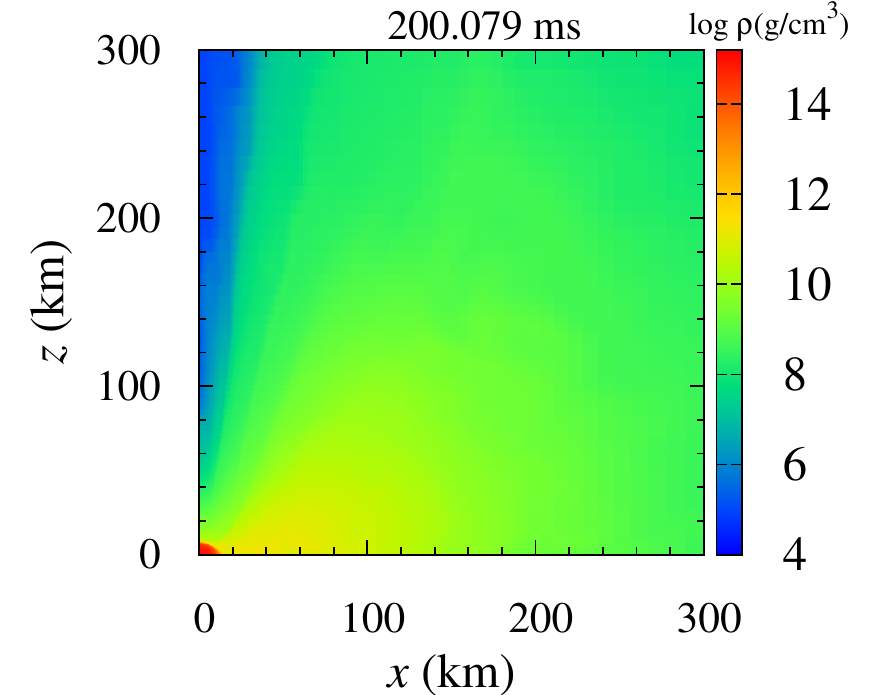} 
\caption{Evolution of density profiles for the model with
  $\alpha_v=0.01$. The upper and middle rows show the early-time
  profiles in $0 \leq x \leq 100$\,km and $0 \leq z \leq 100$\,km
  while the bottom rows show the late-phase profiles in $0 \leq x \leq
  300$\,km and $0 \leq z \leq 300$\,km. Time is shown in the upper
  region of each plot.  A plot for a wider region of the bottom right
  panel is found in the middle-middle panel of Fig.~\ref{fig5}.
\label{fig3}}
\end{center}
\end{figure*}

\begin{figure}[th]
\begin{center}
\includegraphics[width=88mm]{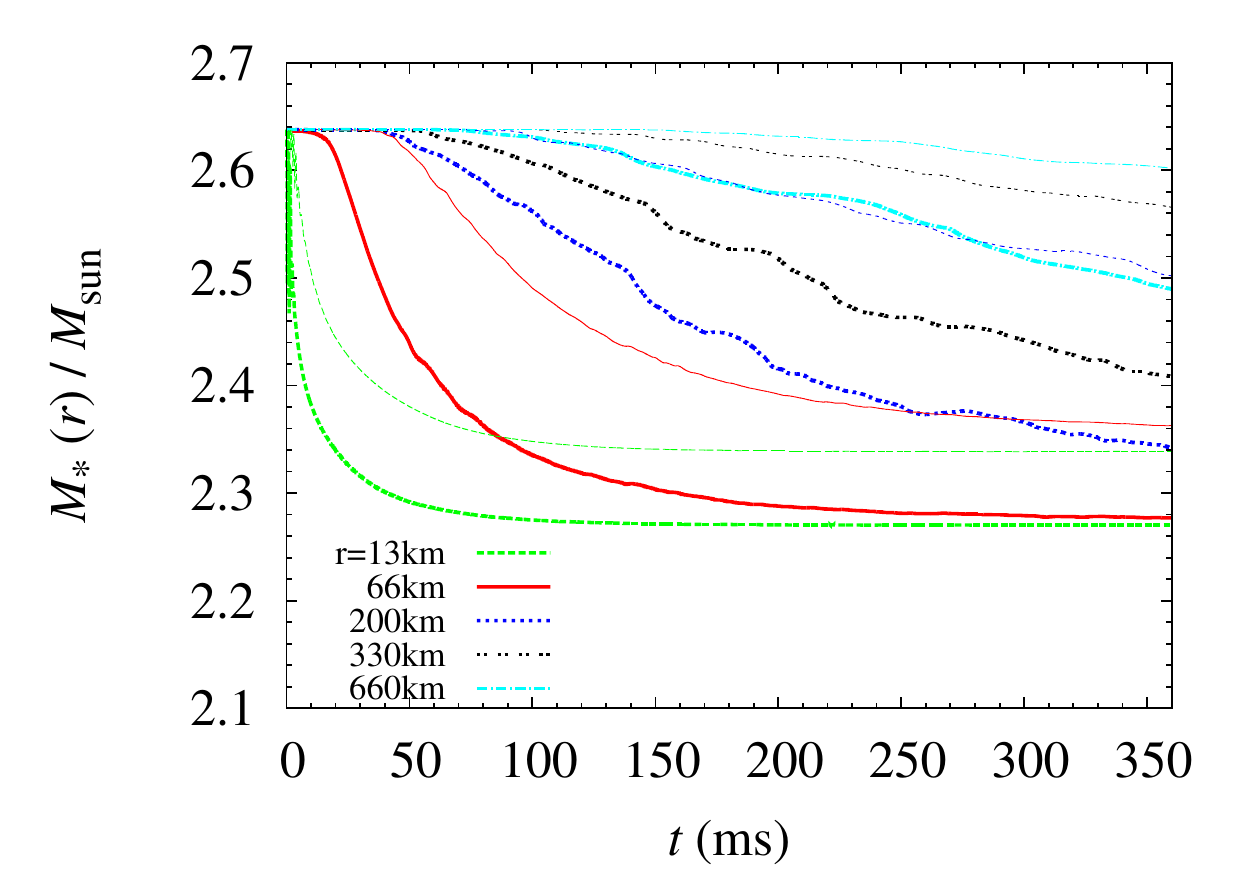}
\caption{The rest mass contained in given radii ($r=13$, 66, 200, 330,
  and 660\,km) as functions of time for $\alpha_v=0.01$ (thin curves)
  and 0.03 (thick curves).
 \label{fig7}}
\end{center}
\end{figure}

\begin{figure}[t]
\begin{center}
\includegraphics[width=84mm]{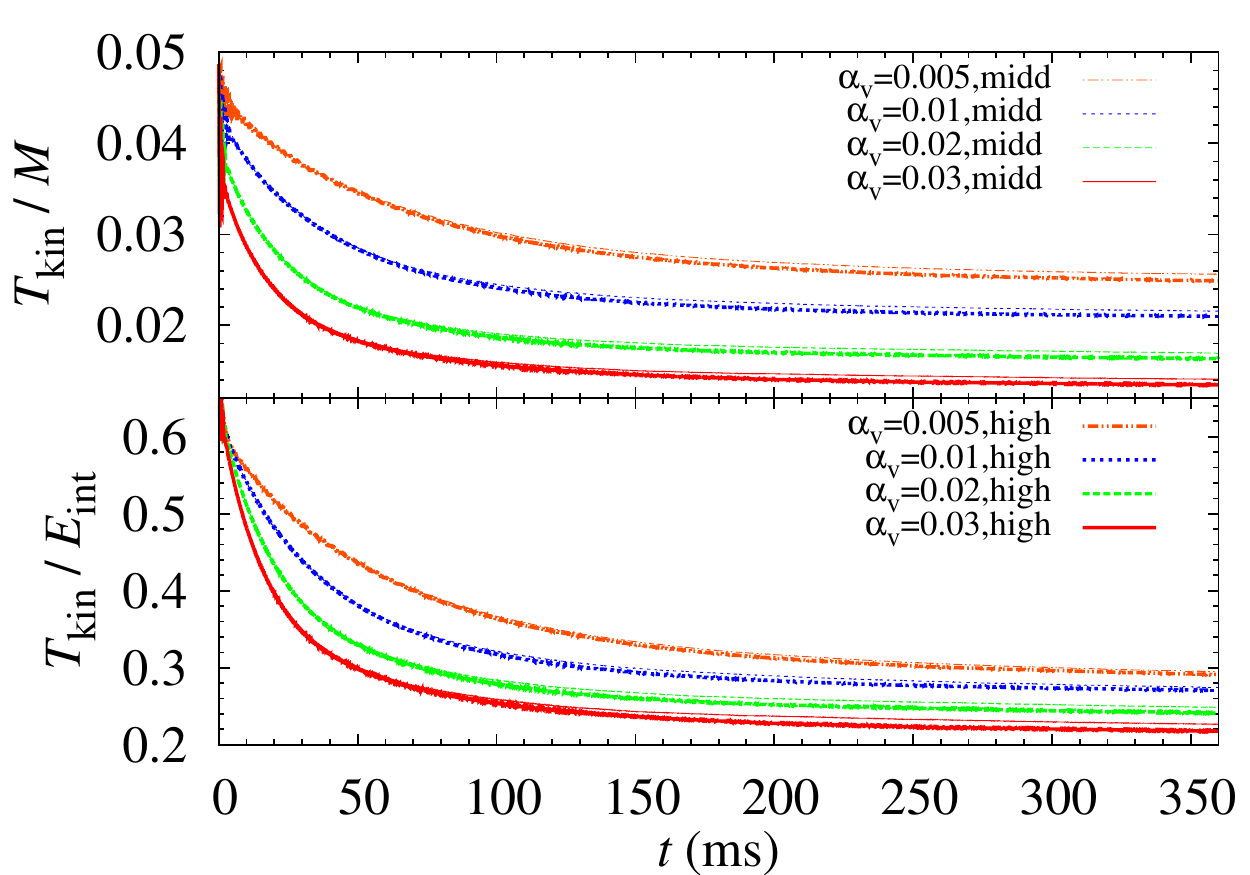}
\caption{Evolution of total kinetic energy (upper panel) and ratio of
  the total kinetic energy to total internal energy (lower panel) for
  $\alpha_v=0.005$--0.03 and for the middle and high grid resolutions.
  Note that the initial values of $T_{\rm kin}/M$ and $T_{\rm
    kin}/E_{\rm int}$ are $\approx 0.048$ and $0.77$, respectively:
  These values significantly decrease in the initial relaxation phase
  for $\sim 10$\,ms during which the neutron star expands and the
  expansion fraction is larger for the larger values of $\alpha_v
  \zeta$.
\label{fig4}}
\end{center}
\end{figure}

\begin{figure*}[thb]
\begin{center}
\includegraphics[width=56mm]{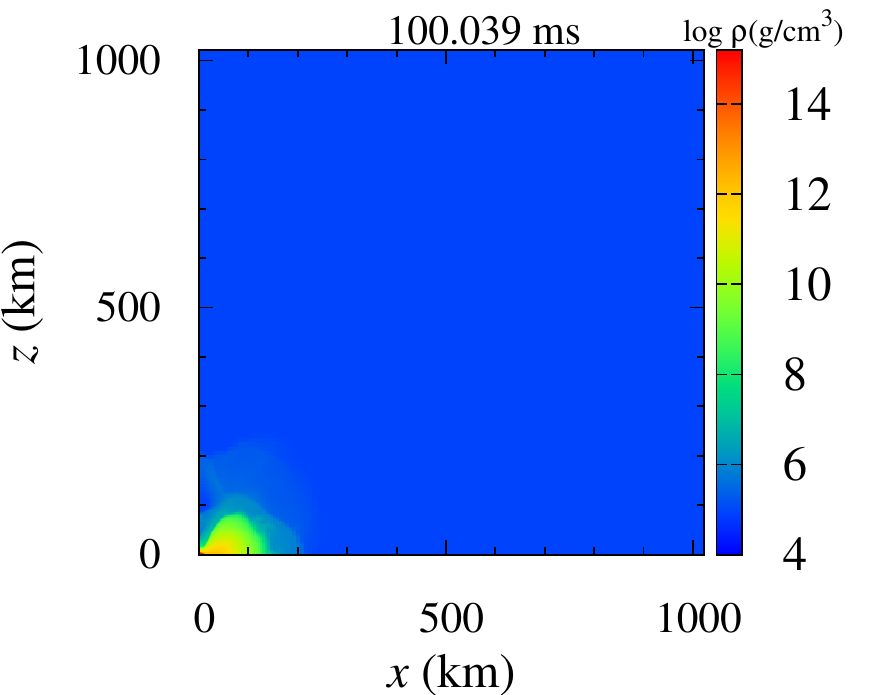}
\includegraphics[width=56mm]{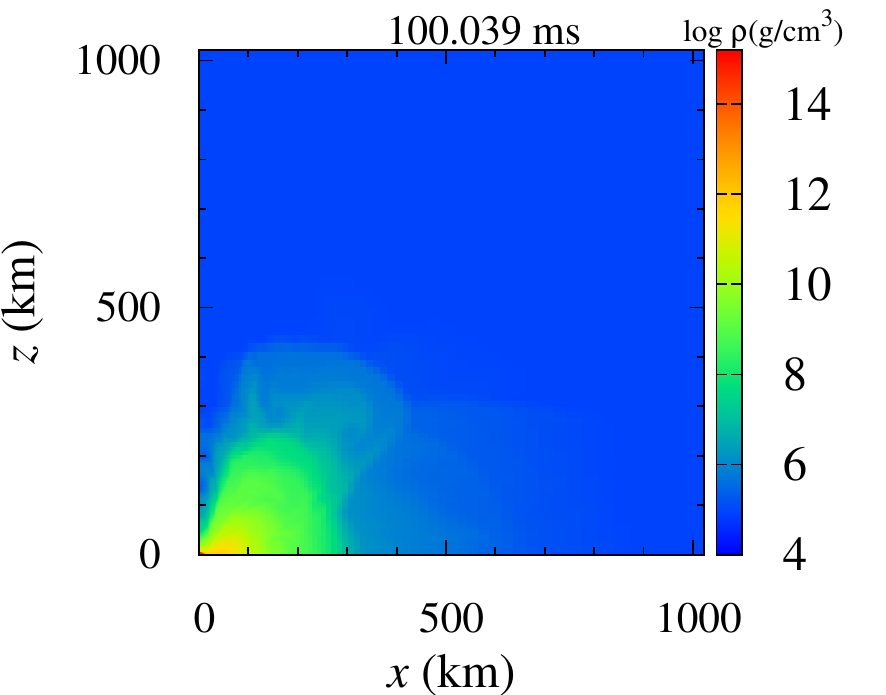}
\includegraphics[width=56mm]{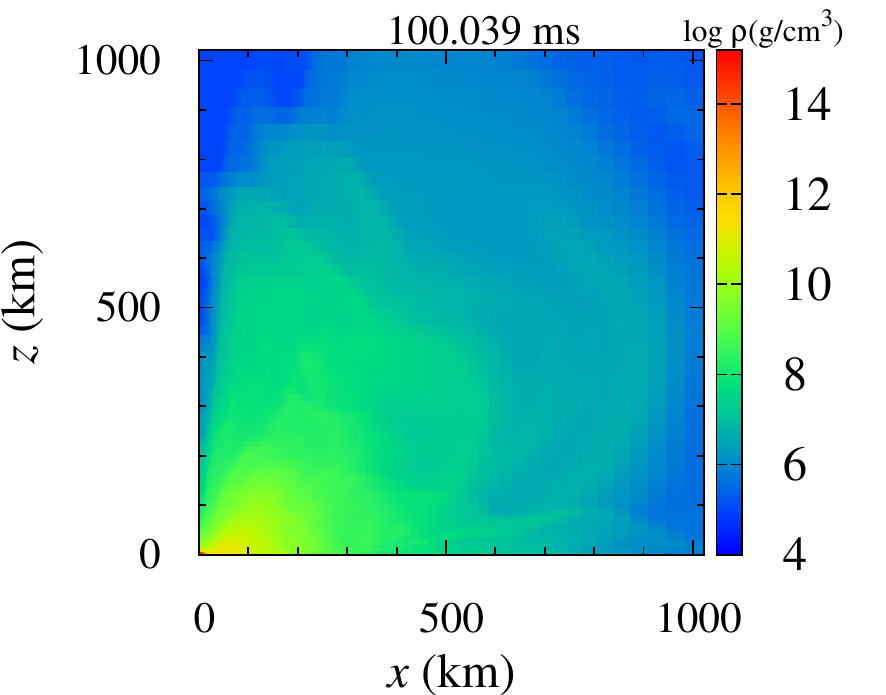} \\
\vspace{1mm}
\includegraphics[width=56mm]{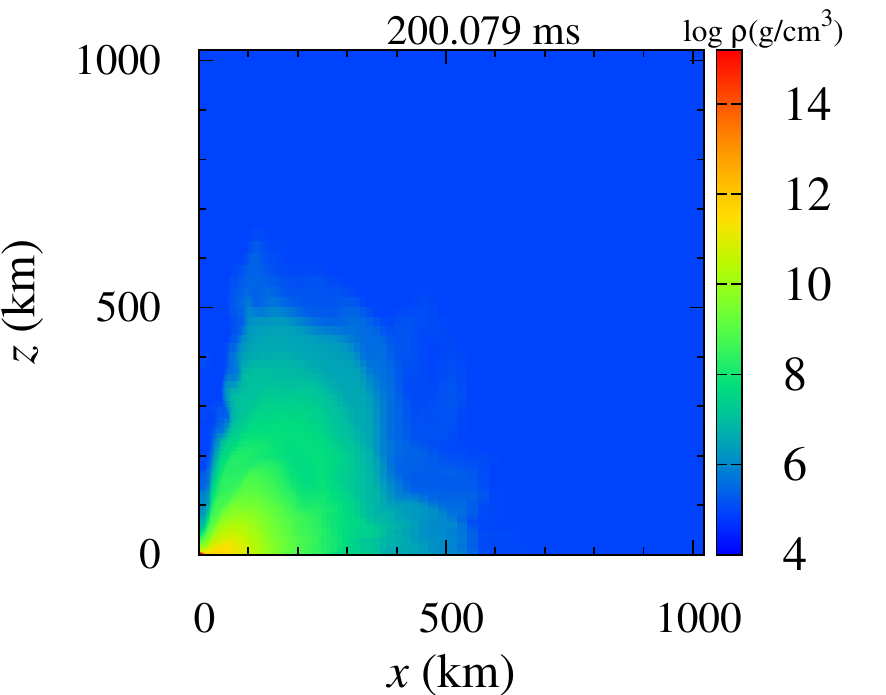}
\includegraphics[width=56mm]{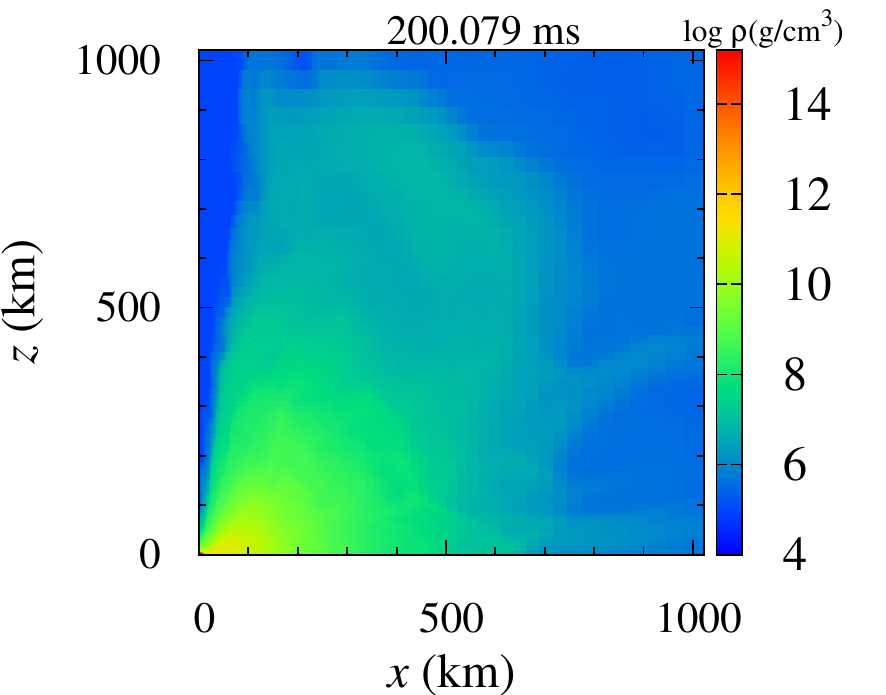}
\includegraphics[width=56mm]{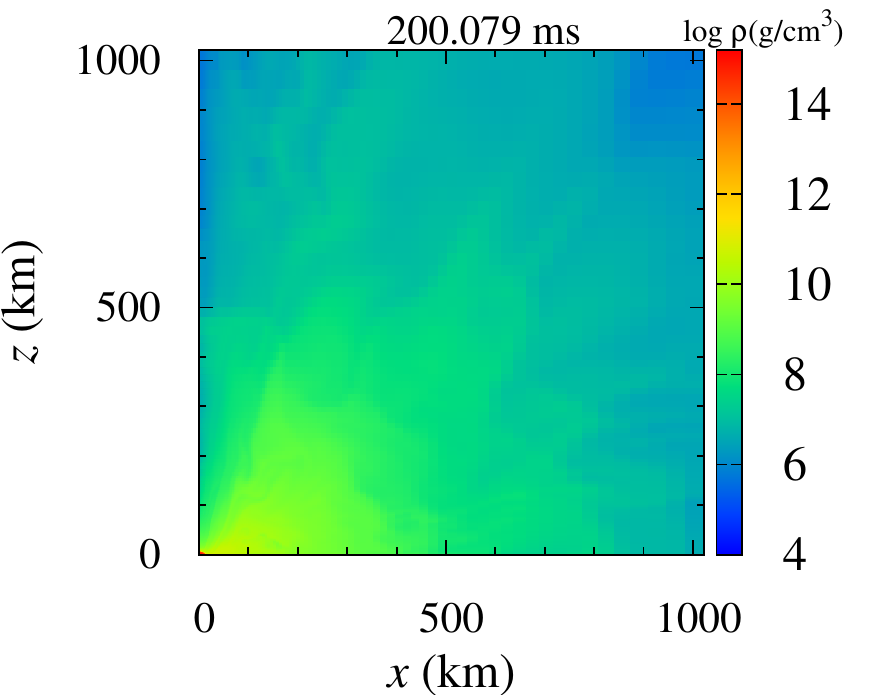} \\
\includegraphics[width=56mm]{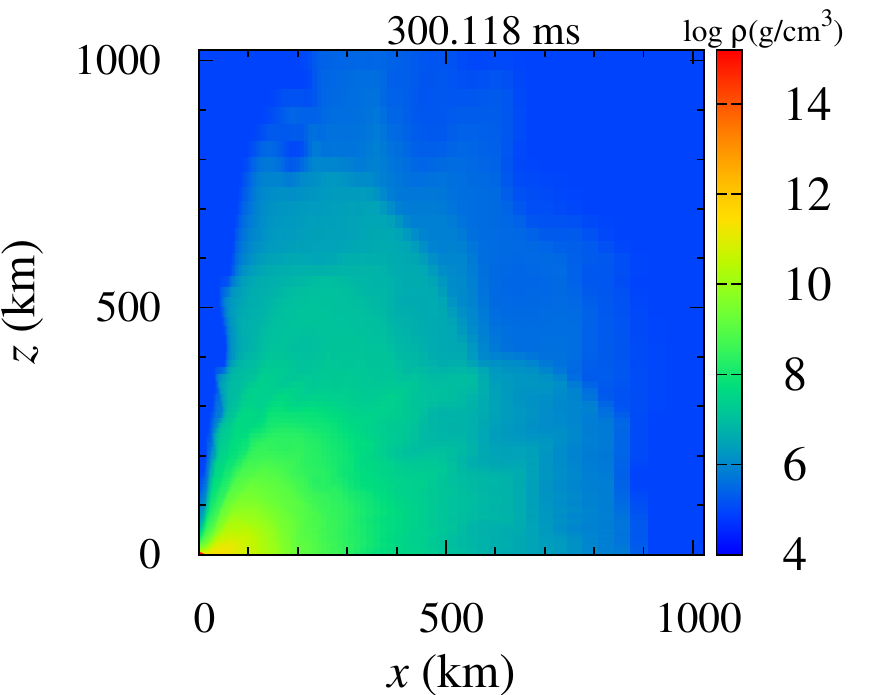}
\includegraphics[width=56mm]{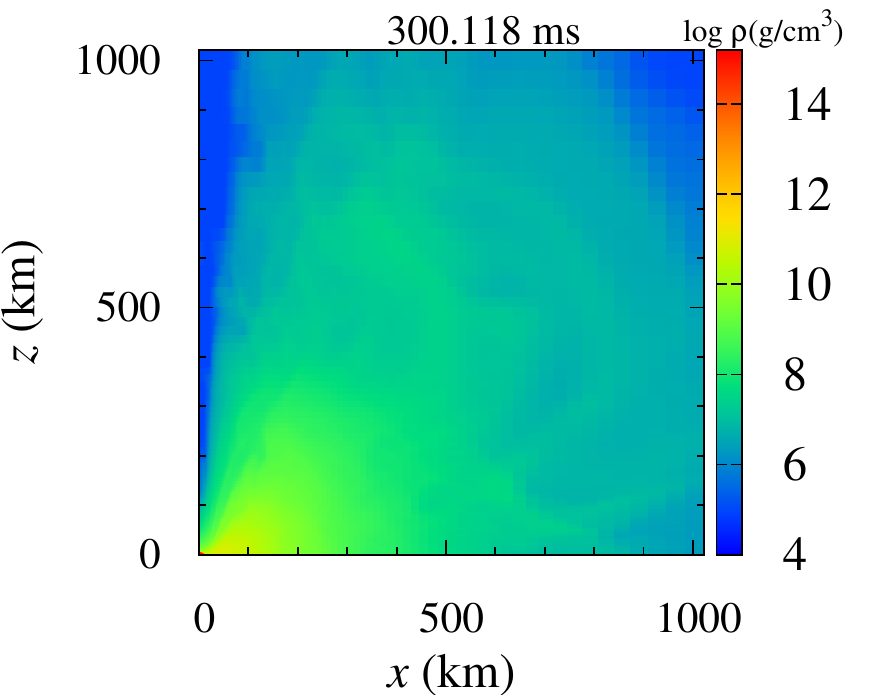}
\includegraphics[width=56mm]{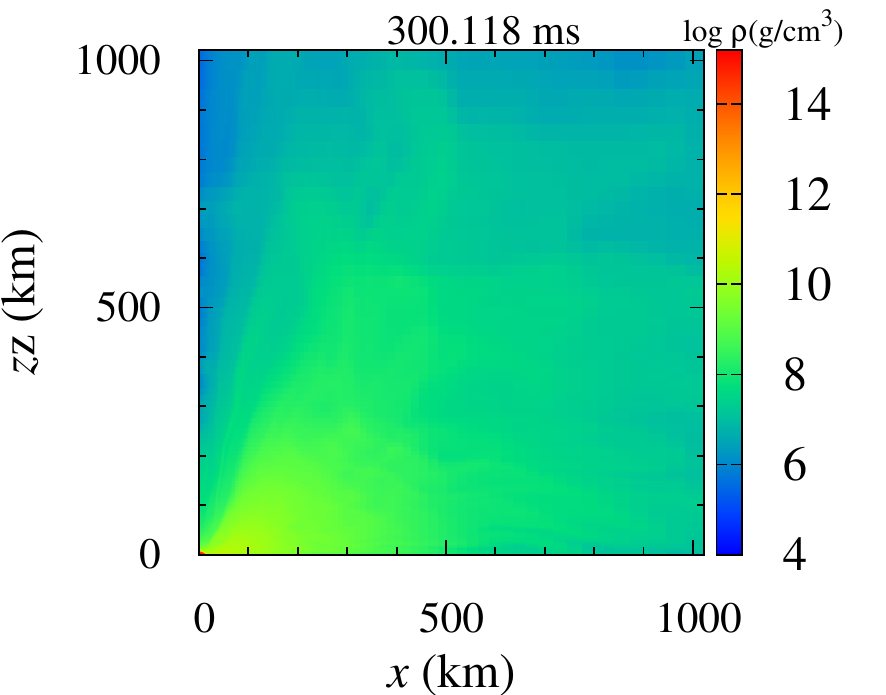} \\
\caption{The density profile for the models with $\alpha_v=0.005$ (left
  column), 0.01 (middle column), and 0.03 (right column) at $t \approx
  100$\,ms (upper), 200\,ms (middle), and 300\,ms (lower).
\label{fig5}}
\end{center}
\end{figure*}

\begin{figure}[t]
\begin{center}
\includegraphics[width=84mm]{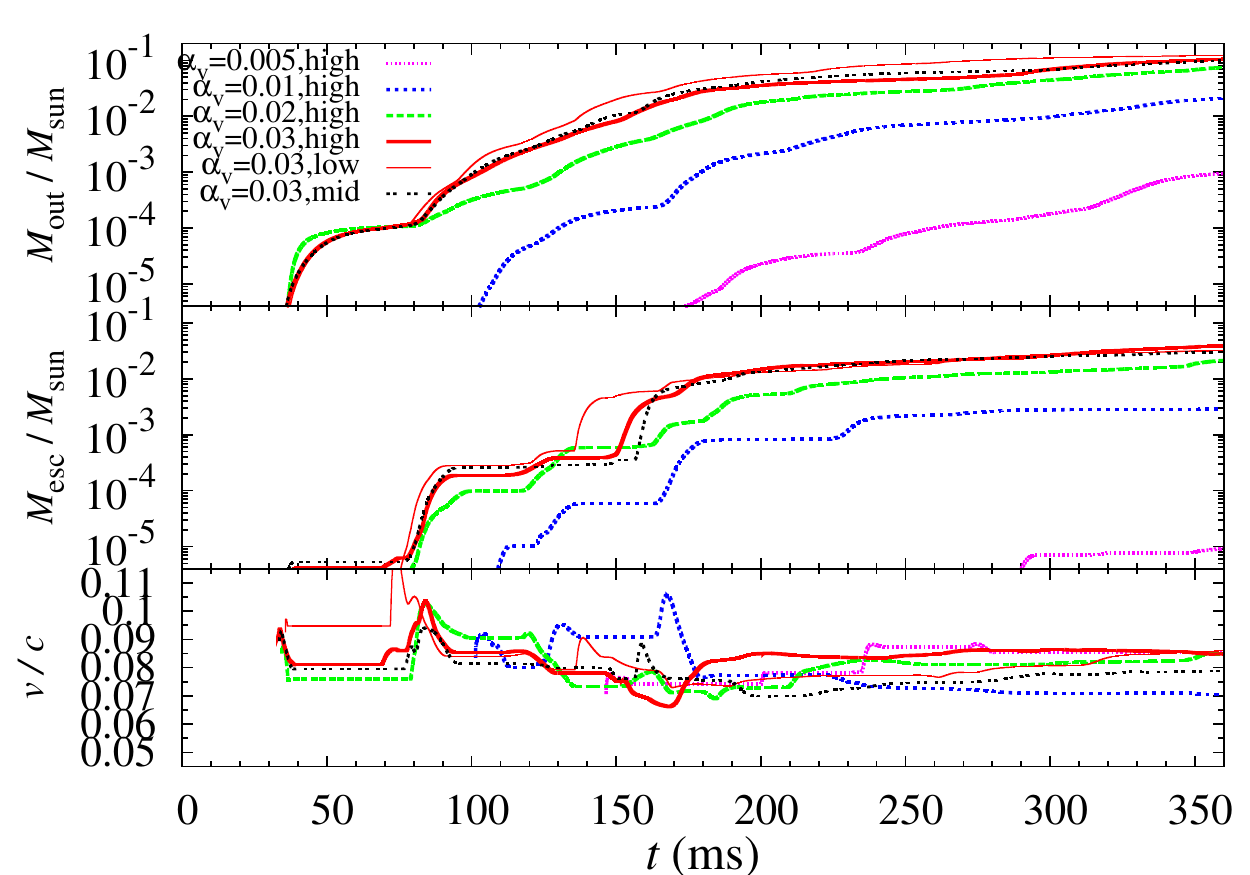}
\caption{Upper and middle panels: The outflow and ejecta mass as
  functions of time for several models with different values of
  $\alpha_v$.  Bottom panel: Averaged velocity of the ejecta component
  as a function of time.  For $\alpha_v=0.005$, 0.01, and 0.02, the
  results in the high-resolution runs are plotted while for
  $\alpha_v=0.03$, the results with three grid resolutions are
  plotted.
 \label{fig6}}
\end{center}
\end{figure}

First, we summarize the typical viscous evolution process of our
differentially rotating neutron star model, paying attention to 
the case of $\alpha_v=0.01$.

In the very early stage of its evolution, angular momentum in the
central region of the neutron star is efficiently transported outward
in $\sim 10 P_0$ where $P_0=2\pi/\Omega_0$: the rotation period along
the rotation axis. As a result, the initially differentially rotation
state changes to an approximately rigidly rotation state for $x \alt
R_e$: see Fig.~\ref{fig1} for the evolution of the profile of the
angular velocity $\Omega$ along the $x$-axis. In the present model,
the angular velocity in the (approximately) rigidly rotating state is
only slightly smaller than the Kepler velocity defined by
$\sqrt{M/R_e^3}$ which is $9.0 \times 10^3$\,rad/s initially. (For
larger values of $\alpha_v$, the relaxed angular velocity is slightly
smaller: see the right panel of Fig.~\ref{fig1}.)  After a nearly
rigidly rotating state is achieved, the effect of the angular momentum
transport inside the neutron star becomes weak and its angular
velocity decreases only slowly as a result of the outward
angular-momentum transport induced by the viscous effect that occurs
in the outer region of the neutron star: see the right panel of
Fig.~\ref{fig1}. Since the angular velocity in the vicinity of the
rotation axis is monotonically and steeply reduced in a few ms, we
plot only the subsequent evolution in Fig.~\ref{fig1}.  This figure
shows that although the decrease time scale of the angular velocity is
quite long, it is certainly reduced in a time scale of $\sim
10^2\,{\rm ms}$. This is due to the presence of the dense envelope and
dense torus surrounding the neutron star to which the angular momentum
is gradually transported from the main neutron-star body. Thus, the
time scale of $\sim 10^2$\,ms is determined by the evolution time
scale of the torus (see below).  Figure~\ref{fig1} also shows that the
numerical results for the long-term evolution of the neutron star
depend very weakly on the grid resolution irrespective of the values
of $\alpha_v$.

Figure~\ref{fig2} displays the evolution of density profiles on the
equatorial plane for $\alpha_v=0.01$ and 0.03. Figure~\ref{fig3} also
displays the evolution of density profiles on the $x$-$z$ plane for
$\alpha_v=0.01$. These figures show that in a short time scale after
the onset of the simulations, dense tori with the maximum density
$\sim 10^{12}\,{\rm g/cm^3}$ are formed around the neutron stars. As
Figs.~\ref{fig2} and~\ref{fig3} show, the density of the tori
subsequently decreases with time due to a long-term viscous process.
Specifically, matter expands outward by the viscous heating and
angular-momentum transport (see Fig.~\ref{fig3} and discussion
below). These figures indicate that the viscous braking of the
neutron-star rotation should continue as long as the dense envelope
and torus surrounding it presents (for a time scale of $O(100$\,ms)).
The right panel of Fig.~\ref{fig1} also shows that the spin-down rate
of the neutron star depends only weakly on the grid resolution.

Because of the long-term angular-momentum transport, a dense and
massive torus surrounding the central neutron star is evolved: see
Fig.~\ref{fig3}. The neutron-star mass decreases (torus mass
increases) gradually with time.  Figure~\ref{fig7} displays the rest
mass contained in given radii as functions of time for $\alpha_v=0.01$
(thin curves) and 0.03 (thick curves).  The chosen radii are 13, 66,
200, 330, and 660\,km (dashed, solid, dotted, dot-dot, and dash-dot
curves). Up to $t \sim 100$\,ms, the matter is ejected from the
central neutron star and constitutes a torus, and for $t \agt
100$\,ms, the rest mass of the neutron star is approximately fixed
(see the dashed curves labeled by $r=13$\,km).  At $t \sim 100$\,ms,
the neutron-star rest mass, defined by the rest mass for $r \leq R_e$,
is reduced to 89\% and 85\% of the total rest mass for $\alpha_v=0.01$
and 0.03, respectively. 


The torus mass should depend on the initial profile of the angular
velocity and the compactness of the neutron star.  Since we knew that
tori of such high mass and high density are often formed around the
massive neutron star in the simulations of binary neutron star
mergers~\cite{STU2005,Hotokezaka2013}, in the present work, we chose
the initial condition that could form the object similar to the merger
remnant of binary neutron stars.

After the formation of the torus, the matter in the torus expands
outward. This is found from Fig.~\ref{fig7}: Differences between the
dashed curves of $r=13$\,km and any other curves decrease with time;
e.g, for $\alpha_v=0.03$, by comparing the curves of $r=13$ and 
66\,km, we find that the mass of the inner part of the torus is $\sim
0.1M_\odot$ at $t=50$\,ms and it decreases to $\sim 0.02M_\odot$ at
$t=300$\,ms.

The matter of the torus on the equatorial plane has nearly Keplerian
motion~(see the left panel of Fig.~\ref{fig1}). Thus, in the outer
envelope of the neutron star and in particular in the torus,
differential rotation remains, and hence, viscous angular momentum
transport continuously works in the outer part of the system.
Consequently, the viscous heating plays an important role even after
the neutron star settles to a rigidly rotating state. The time scale
for this process should be much longer than the viscous time scale in
the differentially rotating neutron star, because the values of $R$
and $\nu^{-1}$ are larger in the outer part than the inner part (see
Eq.~(\ref{eq3.5}) for the definition of the viscous time scale,
$t_{\rm vis}$).


Figure~\ref{fig4} plots the evolution of total kinetic energy and the
ratio of the total kinetic energy to total internal energy as
functions of time. Due to the continuous viscous process, the kinetic
energy is dissipated and converted to the internal energy.  For
$\alpha_v=0.005$--0.03, the kinetic energy is decreased by $\sim
50$--70\% until $t\approx 300$\,ms. Here, the dissipation rate of the
kinetic energy is higher for the larger value of $\alpha_v$.  The
ratio of the kinetic to internal energy decreases in a similar manner
to that for the kinetic energy, i.e., the increase rate of the
internal energy is much lower than the decrease rate of the kinetic
energy. Our interpretation for this is that the increase of the
internal energy resulting from the viscous dissipation is consumed by
the adiabatic expansion of the torus, as Fig.~\ref{fig3} indicates
this fact. 

Since no cooling effect except for the adiabatic expansion is taken
into account in this study (although we conservatively include the
shock-heating effect by choosing a small value of $\Gamma$), the
geometrical thickness of the torus is monotonically increased by the
viscous heating. We note that in the presence of a rapidly rotating
neutron star at center (in the absence of a black hole that absorb
matter), torus matter cannot efficiently fall onto the neutron star.
Thus, unless the torus matter is ejected outwards, it continuously
contributes to the viscous heating and resulting increase of its
geometrical thickness.  In reality, the neutrino emission would come
into play for this type of the dense system. The typical neutrino
cooling time scale may be longer than the viscous heating time scale
of $\sim 100$\,ms for the region of the density larger than $\sim
10^{11}\,{\rm g/cm^3}$ because neutrinos are optically thick and
trapped in the torus~\cite{trap}. However, after the torus expands and
its density is decreased, subsequent expansion may be prohibited by
the neutrino cooling. On the other hand, neutrino irradiation may
enhance the torus expansion and mass ejection because the torus and
outer part of the neutron star are quite hot and can be strong
neutrino emitters.  Incorporating the neutrino physics is one of the
issues planned for our future work.
 
As a result of the monotonic increase of the geometrical thickness of
the torus, a funnel structure is eventually formed (see the bottom
panels of Fig.~\ref{fig3}).  Figure~\ref{fig5} displays snapshots of
the density profiles for a wide region of 1000\,km$\times 1000$\.km at
$t \approx 100$, 200, and 300\,ms for $\alpha_v=0.005$ (left), 0.01
(middle), and 0.03 (right), respectively. Vertically expanding matter
is clearly found in this figure.  This is very similar to and
qualitatively the same as the structure found in MHD simulations in
general relativity (e.g., Refs.~\cite{Hawley2006,kiuchi2015}). This
agreement is reasonable because in both cases, viscous or MHD shock
heating enhances the geometrical thickness, and thus, the rotating
matter expands in the vertical direction. This result suggests that
viscous hydrodynamics would capture an important part of the MHD
effects such as shock heating and subsequent torus evolution at least
qualitatively.


Due to the continuous viscous heating in the outer part of the neutron
star and surrounding torus, a part of matter of the torus is outflowed
eventually. Figure~\ref{fig6} displays the rest mass of the outflowed
and ejected matter, $M_{\rm out}$ and $M_{\rm esc}$, and the averaged
velocity of the ejecta as functions of time.  Here, the outflowed
component is estimated from the rest-mass and energy fluxes for a
coordinate sphere at $r=1173$\,km, and if the specific energy for a
fluid component becomes positive, i.e., $u_t < -1$, we specify it as
the ejecta component. The averaged velocity of the ejecta is defined
by $\sqrt{2T_{\rm esc}/M_{\rm esc}}$ where $T_{\rm esc}$ is the
kinetic energy of ejecta; a fraction of $T_{\rm out}$ that satisfies
$u_t<-1$.  We note that the curves for $M_{\rm eje}$ with different
grid resolutions, in general, do not agree well with each other (this
is in particular the case for small values of $\alpha_v$ for which the
ejecta mass is small). Our interpretation for this is that during the
outflow is driven, there are many fluid components which are
marginally unbound with $u_t \approx -1$, and hence, it is not
feasible to accurately specify the ejecta components. However, the
final values of the ejecta mass and kinetic energy depend weakly on
the grid resolutions: These values are determined within a factor of
$\sim 2$.

The upper panel of Fig.~\ref{fig6} shows that irrespective of the
values of $\alpha_v$, a fraction of the matter goes away from the
central region. This is reasonable because geometrical thickness of
the torus surrounding the central neutron star always grows
irrespective of $\alpha_v$ (see Fig.~\ref{fig5}).  The total amount of
the outflowed mass is larger for the larger values of $\alpha_v$ for a
given moment of time, because the viscous heating rate is higher. 

The outflow component comes primarily from the matter originally
located at the torus, as we already mentioned (see Fig.~\ref{fig7}).
Figure~\ref{fig6} indicates that the outflowed mass eventually
converges to a relaxed value for $\alpha_v=0.02$ and 0.03.  This is
because the mass of the torus surrounding the central neutron star
decreases with time as mentioned already. Thus, the final outcome
after the evolution of differentially rotating neutron stars is likely
to be a rigidly rotating neutron star surrounded by a low-density
torus and a widely-spread envelope as Fig.~\ref{fig5} indicates.

Figure~\ref{fig6} shows that for $\alpha_v \geq 0.02$, the total mass
of the ejecta is $\agt 10^{-2}M_\odot$. This value is approximately
equal to or larger than those in the dynamical mass ejection of binary
neutron star mergers, for which the typical ejecta mass is
$10^{-3}$--$10^{-2}M_\odot$~\cite{hotoke13}. Thus, the long-term
viscous mass ejection from the merger remnant may be the dominant
mechanism of the mass ejection (see, e.g.,
Refs.~\cite{Fer,Perego2014,Just2014} for similar suggestions).  On the
other hand, for $\alpha_v=0.005$ and 0.01, the ejecta mass is of order
$10^{-5}$ and $10^{-3}M_\odot$, respectively: Only a small fraction of
the outflow material can be ejecta.  This indicates that to get a
large value of the ejecta mass by the viscous process, an efficient
viscous heating would be necessary (in reality, strong MHD turbulence
would be necessary).

The bottom panel of Fig.~\ref{fig6} shows that the averaged velocity
of the ejecta is $\alt 0.1c$ irrespective of the values of
$\alpha_v$. This is smaller than that for the dynamical mass
ejection~\cite{hotoke13} but the result is consistent with other
viscous hydrodynamics results (see, e.g., Ref.~\cite{Fer}).  As
discussed in Sec.~I, rotating massive neutron stars
surrounded by a massive torus are likely to be canonical outcomes of
the binary neutron star merger. During the binary merger, the matter
would be dynamically ejected, in particular, at the onset of the
merger with the typical averaged velocity $\sim
0.2c$~\cite{hotoke13}. If the remnant massive neutron stars are
long-lived, they may subsequently eject the matter by the viscous
effect. As suggested in this paper, the averaged velocity for it would
be less than half of the velocity of the dynamical ejecta.  Therefore,
the viscous ejecta will never catch up with the dynamical ejecta: The
ejecta are composed of two different components.  In the
binary neutron star mergers, the dynamical ejecta are likely to have a
quasi-spherical or weakly spheroidal morphology~\cite{hotoke13}.
Thus, the viscous ejecta are likely to be surrounded by the dynamical
ejecta.

As described in Refs.~\cite{Li98,kasen13}, in the viscous ejecta as
well as in the dynamical ejecta, r-process nucleosynthesis is likely
to proceed because the ejecta are dense and neutron-rich, and then the
ejecta will emit high-luminosity electromagnetic signals fueled by the
radioactive decay of the unstable r-process heavy elements.  In the
presence of strong viscous wind, there may be two components in the
light curve, while in its absence, the dynamical ejecta would be the
primary source for the electromagnetic signals~\cite{kasen15}.  As
Kasen and his collaborators illustrate, the shape of the light curve
is quite different depending on the presence or the absence of the
viscous wind. Our present result indicates that the viscous ejecta
would be surrounded by the quasi-spherical dynamical ejecta. This
suggests that the emission from the viscous ejecta could be absorbed
by the dynamical ejecta, and then, the absorbed energy could be
reprocessed and power up the emissivity of the dynamical ejecta.

The electromagnetic signal associated with the decay of unstable
r-process elements is one of the most promising electromagnetic
counterparts of the binary neutron star mergers. For the detection of
these electromagnetic counterparts, we need a theoretical prediction
as accurately as possible. The present study suggests that the light
curve of this electromagnetic signal is uncertain due to the
uncertainty of the viscous parameter that determines the ejecta
mass. This implies that for the prediction of the electromagnetic
signals, we have to perform numerical simulations taking into
account a wide variety of the possibilities for the viscous
parameter. Ultimately, we will need to perform a sufficiently
high-resolution MHD simulation with no symmetry that can uniquely
clarify the evolution of the differentially rotating merger remnants
in the first-principle manner.

\section{Summary}

Employing a simplified version of the Israel-Stewart formulation for
general relativistic viscous hydrodynamics that can minimally capture
the effects of the viscous angular momentum transport and the viscous
heating, we successfully performed axisymmetric numerical-relativity
simulations for the evolution of a differentially rotating neutron
star, which results in an approximately rigidly rotating neutron star
surrounded by a massive torus. The detailed evolution process of this
model with a sufficiently high viscous parameter is summarized as
follows.  First, by the outward angular momentum transport process,
the initially differential rotation state is forced to be an
approximately rigid rotation state in the inner region of the neutron
star.  At the same time, the torus with substantial mass is formed due
to the viscous angular momentum transport from the neutron star. The
time scale for this early evolution is quite short $\sim 10$\,ms
(i.e., the viscous time scale of the differentially rotating neutron
star).  The outcome in this stage is similar to the merger remnant of
binary neutron stars.

Subsequently, the torus mass (including envelope surrounding the torus)
increases spending a long time scale $\sim 100$\,ms and eventually
reaches $\sim 0.3$--$0.4M_\odot$ in the present model (this mass
should depend on the initial choice of the models).  After the
formation of the system composed of a (approximately) rigidly rotating
neutron star and a differentially rotating massive torus, the viscous
effect still plays an important role near the outer surface of the
neutron star and in the torus. Due to the subsequent long-term viscous
heating effect there, the thermal pressure of the torus is increased,
and as a result, the geometrical thickness of the torus monotonically
increases. Also, the torus gradually expands along the equatorial
direction because of the viscous angular momentum transport. For a
sufficiently high viscous parameter, eventually, a strong outflow is
driven from the torus. The ejecta mass can reach $\agt 0.01M_\odot$
for $\alpha_v \geq 0.02$ in our model. Therefore, if a viscous process
is efficient for the remnant of binary neutron star mergers, it is
natural to expect the ejecta of large mass that is comparable to or
larger than the mass of the dynamical component ejected during the
merger phase. Since its velocity is likely to be smaller than $0.1c$,
the viscous-driven ejecta will be surrounded by the dynamical ejecta
for which the typical velocity is $\sim 0.2c$.

As we discussed in Sec.~I, the remnants of binary neutron star mergers
are in general differentially rotating objects (typically a massive
neutron star surrounded by a torus), which would be evolved by MHD
turbulence. Thus, in reality, the evolution of the merger remnants
should be determined by the MHD processes, and for clarifying it, we
have to perform a high-resolution non-axisymmetric MHD simulation in
general relativity, for which the resolution has to be higher than the
current best one~\cite{kiuchi}. As we showed in this paper, if the 
effective viscous parameter, $\alpha_v$, is larger than a critical
value, a substantial amount of matter would be ejected from the merger
remnant. Even for the case that $\alpha_v$ is smaller than the
critical value, a large amount of matter could expand to a region far
from the central merger remnant. Thus, the picture for the evolution
of the merger remnant could be significantly different from that in
the absence of the MHD effects. A future high-resolution MHD
simulation is awaited for precisely understanding the evolution
process of the merger remnant. However, in the near future, such
simulations cannot be done because of the restricted computational
resources. The second-best strategy for exploring the mass ejection
process from the merger remnant will be to perform a detailed viscous
hydrodynamics simulation systematically changing the viscous
parameter in a plausible range. 

{\em Note added in proof}: After we submitted this paper, a paper by
David Radice~\cite{radice} was submitted to arXiv. He describes
another viscous hydrodynamics formalism that works well. Although he
focuses only on the case with a small viscous parameter (in the
terminology of alpha viscosity, he focuses only on the cases of
$\alpha_v = O(10^{-3})$ or less), we find that his results agree
qualitatively with our findings.

\begin{acknowledgments}

We thank S. Inutsuka and L. Lehner for helpful discussion on
general-relativistic viscous hydrodynamics.  This work was supported
by Grant-in-Aid for Scientific Research (Grant Nos. 24244028,
15H00782, 15H00783, 15H00836, 15K05077, 16H02183, 16K17706) of
Japanese JSPS and by a post-K computer project (Priority issue No.~9)
of Japanese MEXT.

\end{acknowledgments}

\appendix

\section{Black hole and torus: Test simulation}

In this appendix, we show results of a test simulation for the system
composed of a black hole and a massive torus following the request by
our referee who asks us to demonstrate more evidence that our
formalism is capable of performing a long-term viscous hydrodynamics
simulation. The purpose of this appendix is to demonstrate that our
formalism indeed enables to perform simulations for strongly
self-gravitating systems. More detailed study for the black hole-torus
systems will be presented in a future work.

\begin{figure}[t]
\begin{center}
\includegraphics[width=85mm]{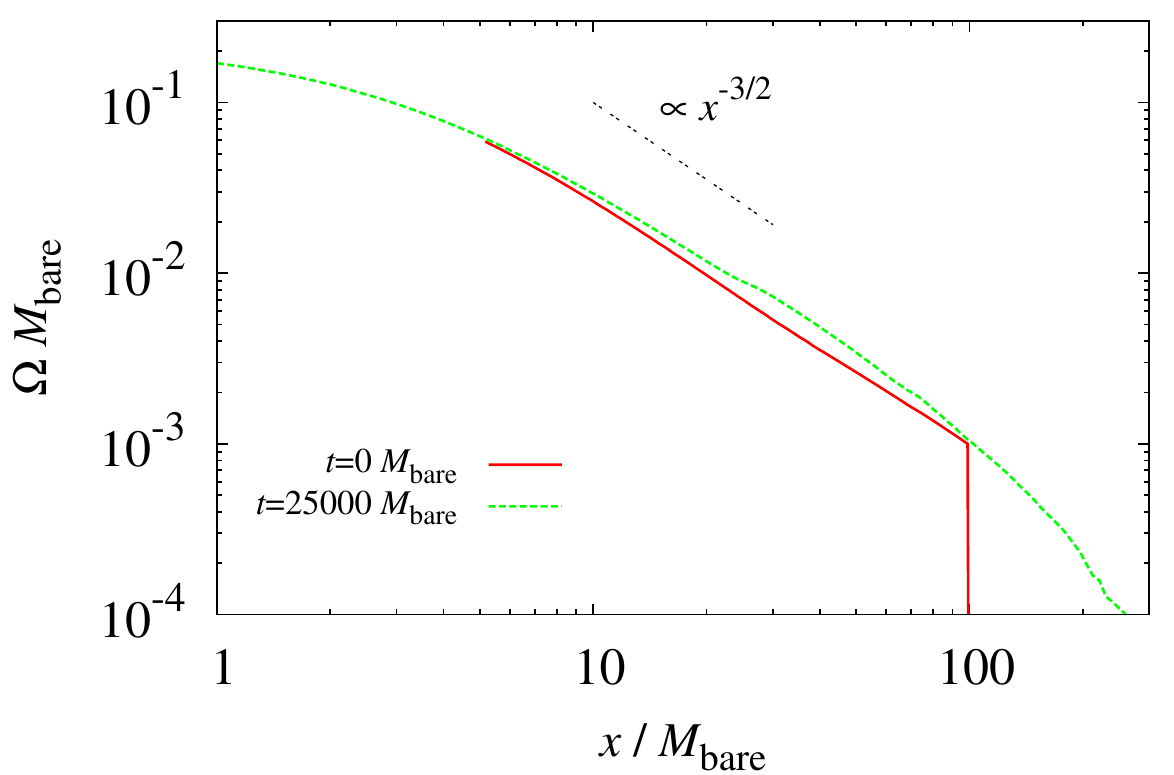}
\caption{Profiles of $\Omega$ as a function of the cylindrical radius
  on the equatorial plane for the torus surrounding the black hole at
  $t=0$ and $t\approx 20560M_{\rm bare}$.  The dot-dot line denotes
  the inclination of $x^{-3/2}$. The angular velocity profiles are
  only slightly modified during the evolution and they appear to be
  always close to the Keplerian one.
  \label{fig9a}}
\end{center}
\end{figure}

\begin{figure*}[t]
\begin{center}
\includegraphics[width=56mm]{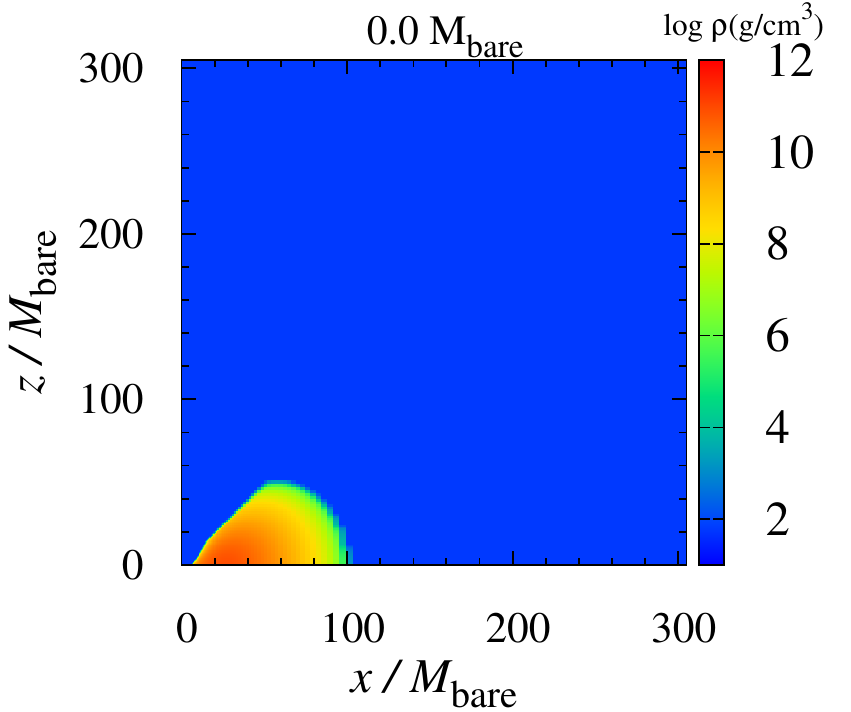}
\includegraphics[width=56mm]{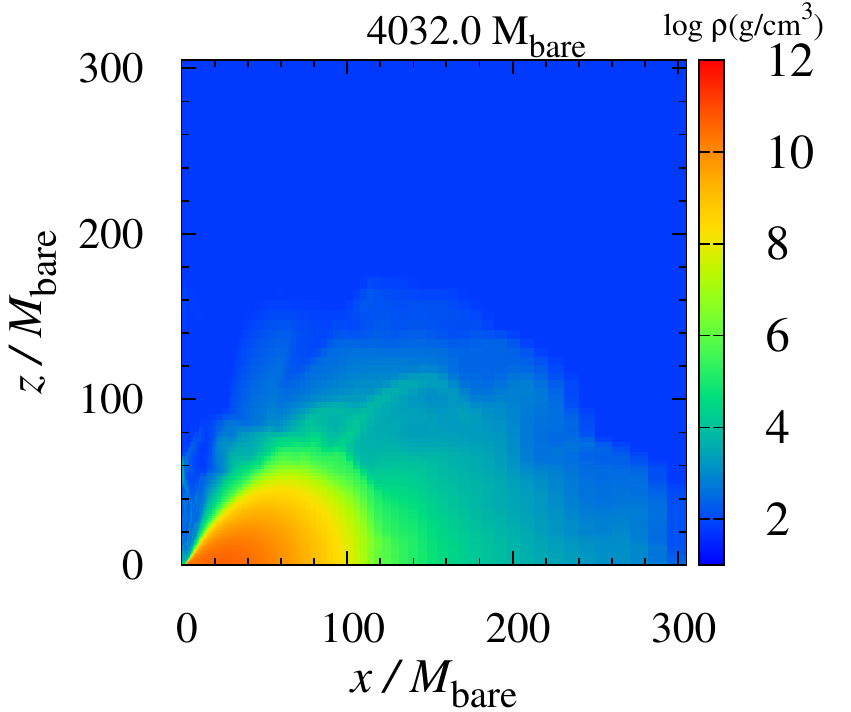}
\includegraphics[width=56mm]{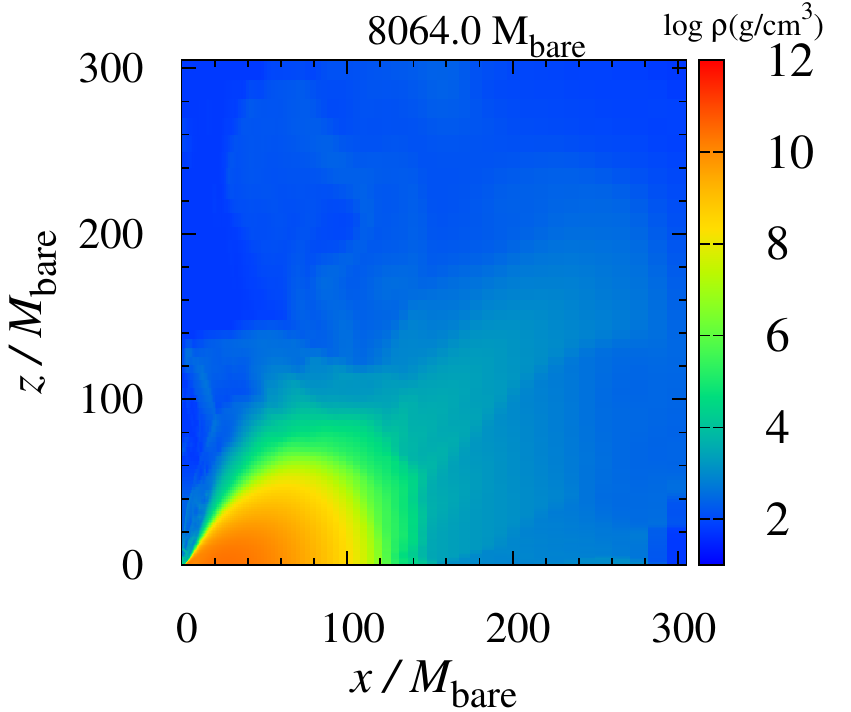} \\
\vspace{0.2cm}
\includegraphics[width=56mm]{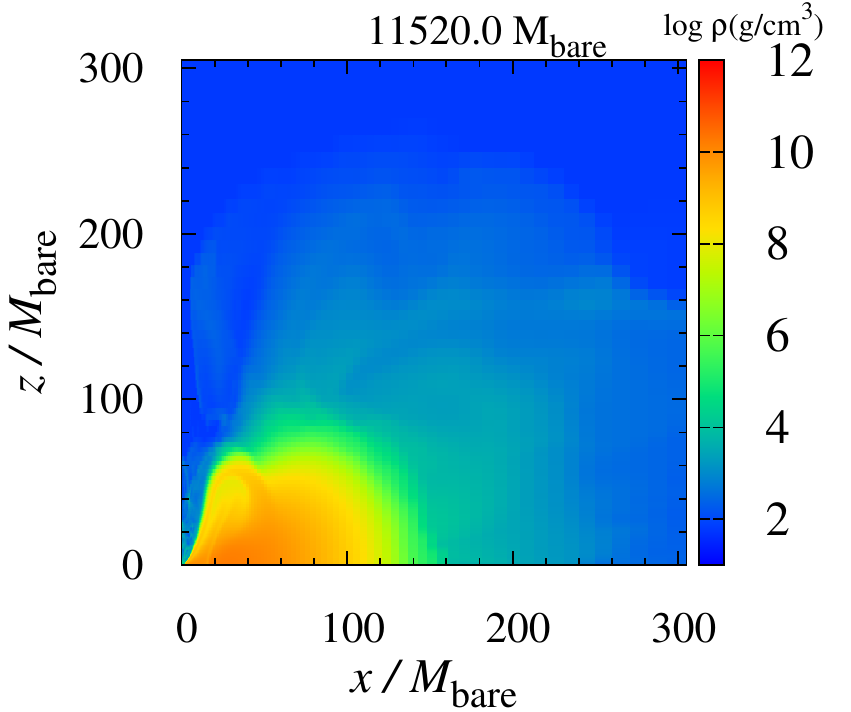}
\includegraphics[width=56mm]{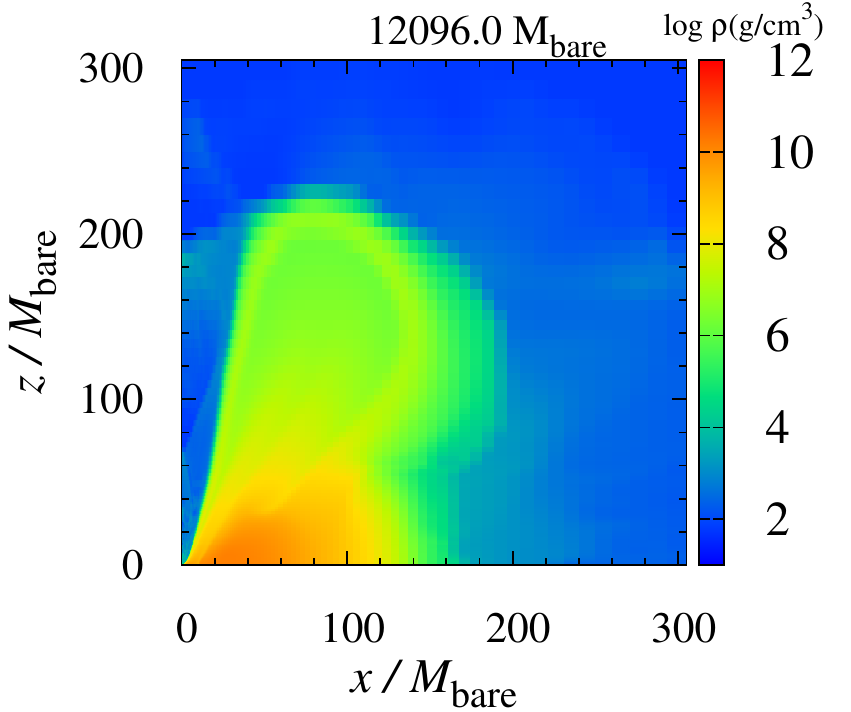}
\includegraphics[width=56mm]{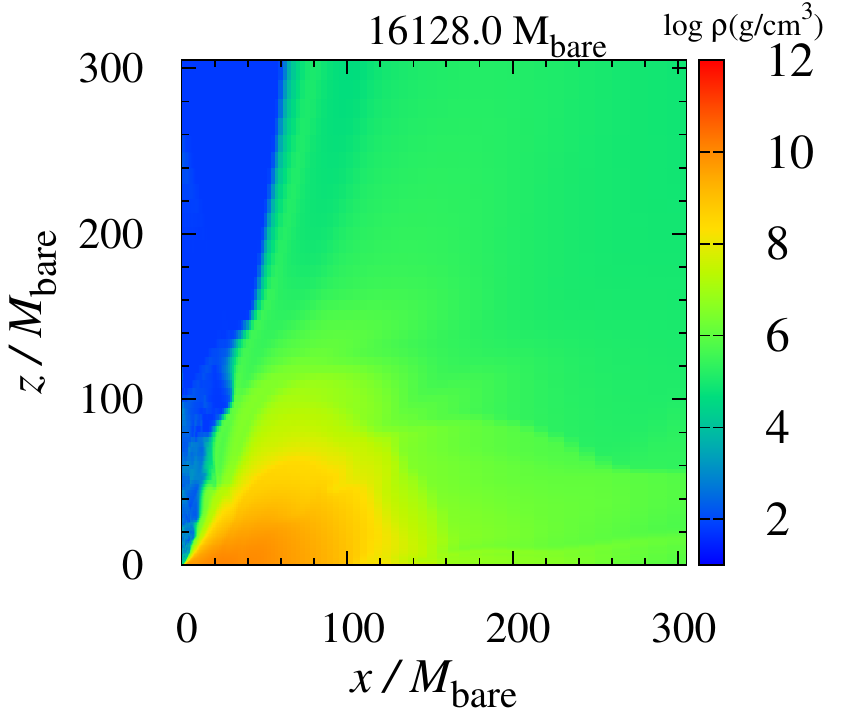}
\caption{The same as Fig.~\ref{fig4} but for the evolution of density
  profiles for the system of a black hole and a massive torus.
  Time and spatial coordinates are show in units of $M_{\rm bare}$. 
\label{fig9b}}
\end{center}
\end{figure*}

For this simulation, we prepare an equilibrium state composed of a
black hole and a massive torus as the initial condition using the
method of Ref.~\cite{shibata07}. For this equilibrium state, we employ
a non-rotating black hole with the puncture mass $M_{\rm bare}$
surrounded by a massive torus with the rest mass $2.356M_{\rm bare}$.
The initial black-hole mass measured by the area of the black-hole
horizon is $M_0=1.072M_{\rm bare}$. We note that the initial black
hole mass is slightly different from $M_{\rm bare}$ because of the
presence of the massive torus. The torus is modeled by the
$\Gamma=4/3$ polytropic equation of state and during the simulation,
we employ $P=\rho\varep/3$ as the equation of state.  Following
Ref.~\cite{kiuchi11}, we determine the specific angular momentum of
the torus by providing the relation of $j=j(\Omega) \propto
\Omega^{-1/4}$ where $j$ and $\Omega$ are the specific angular
momentum and angular velocity, respectively.  Note that for $j \propto
\Omega^{-b}$ with $b \rightarrow 1/3$, the velocity profile approaches
the Keplerian. With our choice of $b=1/4$, the velocity profile looks
close to the Keplerian (see Fig.~\ref{fig9a}).  The inner and outer
edges of the torus are set to be $5M_{\rm bare}$ and $100M_{\rm bare}$
(see the first panel of Fig.~\ref{fig9b}) .  In the following, we
employ a unit in which $M_{\rm bare}=10M_\odot$ for showing the
density.

In reality, the system with such massive torus would be unstable to
non-axisymmetric instability like the Papaloizou-Pringle
instability~\cite{PP84}, even though the angular velocity profile is
far from that of the $j=$const law.  The purpose of this test
simulation is to confirm that our viscous hydrodynamics formalism
enables us to perform a long-term stable simulation for this
self-gravitating system. Hence, disregarding the non-axisymmetric
instability, we perform an axisymmetric simulation.

\begin{figure}[t]
\begin{center}
\includegraphics[width=85mm]{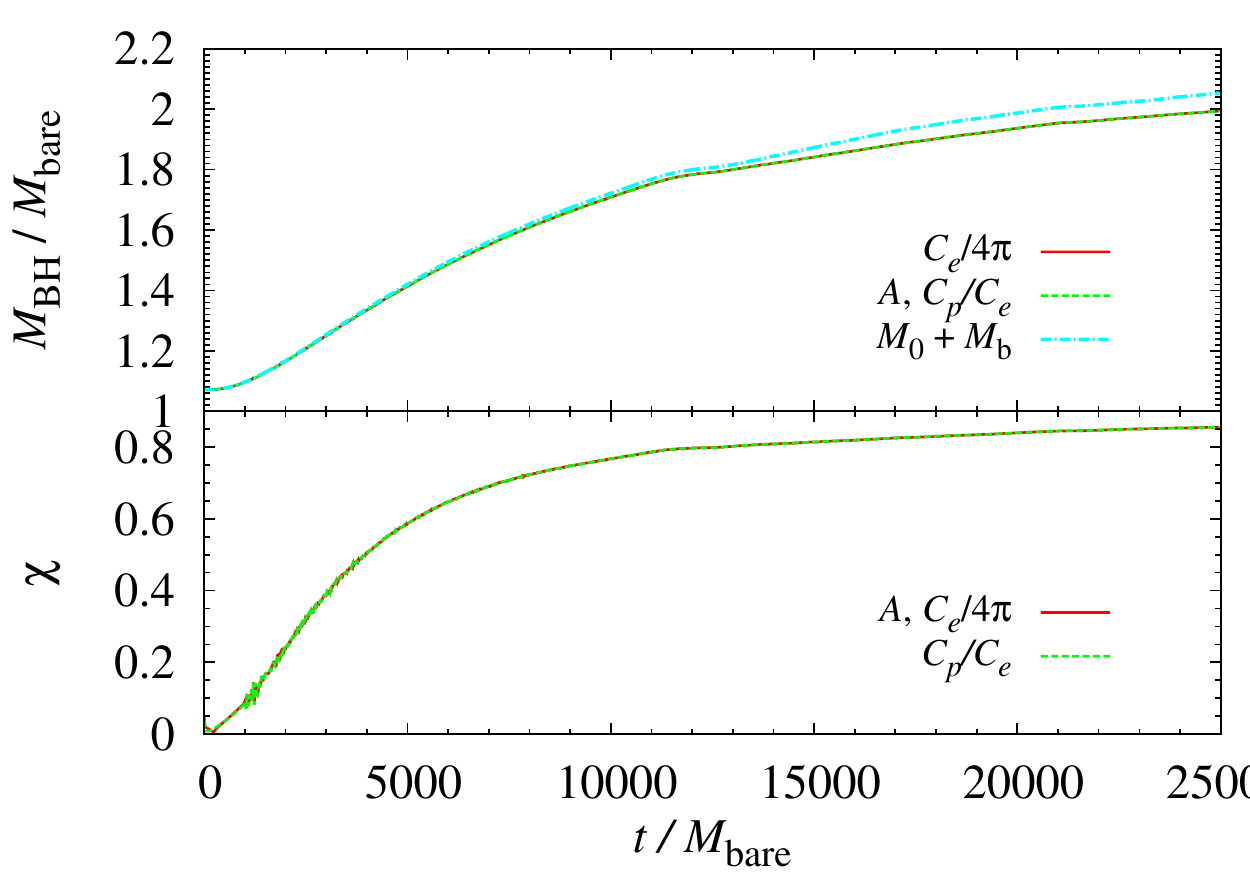}
\caption{Evolution of the mass (upper) and dimensionless spin
  parameter (lower) of the black hole. The mass is determined by
  analyzing $C_e/4\pi$ (solid curve) and $A_{\rm AH}$ together with
  the spin determined by $C_p/C_e$ (dashed curve). The dotted curve
  denotes $M_0 + M_{\rm b}$.  The dimensionless spin is determined
  from $A_{\rm AH}$ together with the mass determined by $C_e/4\pi$
  (solid curve) and from $C_p/C_e$ (dotted curve).
  \label{fig10}}
\end{center}
\end{figure}

In this test simulation, we set $\nu=\alpha_v c_s^2 /\Omega_i$ where 
we choose $\Omega_i=6^{-3/2}M_{\rm bare}^{-1}$, $\alpha_v=0.1$,
$\zeta=3\Omega_i$: We employ a high value of $\alpha_v$ to accelerate
the evolution.  The set up of the computational domain is as follows:
$\Delta x_0=0.03M_{\rm bare}$, $x_{\rm in}=1.2M_{\rm bare}$, and
$f=1.01$ (see Sec III A for these quantities). The outer boundary
along each axis is located at $\approx 555 M_{\rm bare}$.

During the viscous hydrodynamics process, angular momentum transport
actively works in the torus, and as a result, a part of the matter of
the torus falls into the black hole. Then, the mass and spin of the
black hole increase monotonically until the spin parameter reaches a
sufficiently high value. After the high-spin state is reached, the
evolution speed of the black hole is decelerated because the specific
angular momentum at the innermost stable circular orbit around the 
high-spin black hole becomes lower than the values of most of the
torus matter and the infalling of the matter into the black hole is
suppressed.  In this test simulation, we follow the evolution of the
system until the dimensionless spin, $\chi$, is relaxed to be $\approx
0.85$ (see Fig.~\ref{fig10}).

For the analysis of this process, we have to determine the mass and
spin of the black hole. Using the methods described in 
Ref.~\cite{shibata07}, we analyze the quantities of the apparent
horizons of the black hole.  First, assuming that the black hole has
the same properties as Kerr black holes even in the case that it is
surrounded by the matter, the mass of the black hole is determined by
\beqn
M_C :={C_e \over 4\pi}~~{\rm and}~~M_{\rm BH}={2M_{\rm irr} \over
  1+\sqrt{1-\chi^2}}, \label{A1}
\eeqn
where $C_e$ is the equatorial circumferential length of horizons and
$M_{\rm irr}$ is the irreducible mass of the black hole which is
determined from the area of apparent horizons by $\sqrt{A_{\rm
    AH}/16\pi}$.  We will describe the method to determine $\chi$ in
the next paragraph. Note that all the geometrical quantities are
determined for the apparent horizons. We remark that for Kerr black
holes, $M_C=M_{\rm BH}$ is satisfied.  We also approximately estimate
the mass of the black hole by summing up the total rest mass of the
matter swallowed by the black hole, $M_{\rm b}$, and the initial black
hole mass, $M_0$. This approximate mass of the black hole is referred
to as $M_0+M_{\rm b}$ in the following. We note that the energy of the
matter swallowed by the black hole would be smaller than $M_{\rm
  b}$ because of the presence of the gravitational binding energy, and
hence, $M_0 + M_{\rm b}$ would slightly overestimate the black hole
mass (as shown in Fig.~\ref{fig10}).

We determine $\chi$ by two methods. In the first method, we measure
$C_p/C_e$ which is a monotonically decreasing function of $\chi$. Here
$C_p$ is the meridian circumferential length of horizons.  Using the
value of $\chi$ determined by this method, we subsequently determine
$M_{\rm BH}$ shown in Eq.~(\ref{A1}). In the second method, we use the
relation of $M_C=M_{\rm BH}$ for determining the value of $\chi$.

Figure~\ref{fig9b} displays the evolution of the density profile of
the torus. Due to the angular momentum transport inside the torus, a
part of the matter falls into the black hole and another part of the
matter expands outwards (second and third panels of Fig.~\ref{fig9b}).
By the long-term viscous heating effect, the inner part of the torus
is heated up significantly, in particular, a high-spin state with
$\chi \agt 0.8$ is reached (fourth panel of Fig.~\ref{fig9b}), and
then, it expands to a vertical direction. By this outflow, a part of
the torus matter is ejected from the system (fifth panel of
Fig.~\ref{fig9b}).  Eventually, a funnel structure is formed along the
rotation axis of the black hole (sixth panel of Fig.~\ref{fig9b}). In
this final stage of the evolution, the dimensionless black hole spin
is $\approx 0.85$ (see Fig.~\ref{fig10}).

Figure~\ref{fig10} plots the evolution of the mass and dimensionless
spin of the black hole.  We note that for $\chi \ll 0.1$, the accuracy
for the determination of $\chi$ is not very good because the values of
$C_p/C_e$ and $M_{\rm irr}/(C_e/4\pi)$ are close to unity irrespective
of the value of $\chi$.

Besides such an early phase of the evolution, it appears that the
quantities of the black hole are determined accurately because two
independent methods for determining the mass and spin give
approximately the same values. In addition, $M_0+M_{\rm b}$ agrees
approximately with the black hole mass determined by two methods.  It
is also reasonable that $M_0 + M_{\rm b}$ is slightly larger than
$M_C$ and $M_{\rm BH}$.

It is found that by the viscous accretion process, the system
eventually relaxes to a system of a rapidly rotating black hole
surrounded by a geometrically thick accretion torus. Such outcome
is often found in general relativistic MHD simulations (e.g.,
Refs.~\cite{Hawley2006,kiuchi2015}). Our viscous hydrodynamics
simulation captures such feature. 



\end{document}